\documentclass[sigconf,natbib=true]{acmart}

\usepackage{amsmath,amsfonts,amsthm}
\usepackage{graphicx}
\usepackage{algorithm}
\usepackage{algpseudocode}
\usepackage{textcomp}
\usepackage{xcolor}
\usepackage{enumitem}
\usepackage{mathtools}
\usepackage{balance}

\usepackage{booktabs} 
\usepackage{subcaption} 

\usepackage{svg}
\usepackage{fancyhdr}
\usepackage{tabularx}
\usepackage{multirow}
\usepackage{multicol}
\usepackage{mdwlist}
\usepackage{indentfirst}
\usepackage{physics}
\usepackage{tablefootnote}
\usepackage{footnote}
\usepackage[bottom]{footmisc}
\usepackage{hyperref}
\usepackage{makecell}
\usepackage{subcaption}
\usepackage{xspace}
\usepackage{threeparttable}
\usepackage{diagbox}	
\usepackage{bbding} 
\usepackage[hyphenbreaks]{breakurl}
\xspaceaddexceptions{]\}}
\newcommand{\method}{\textsc{LeapRec}\xspace}

\newcommand{\methodfullul}{Calibration-Disentangled \underline{Lea}rning and Relevance-\underline{P}rioritized Reranking\xspace}

\newtheoremstyle{problemstyle}  
        {3pt}                                               
        {3pt}                                               
        {\normalfont\itshape}                               
        {}                                                  
        {\bfseries}                 
        {\normalfont\bfseries:}         
        {.5em}                                          
        {}                                                  
\theoremstyle{problemstyle}
\newtheorem{property}{Property}
\newtheorem{definition}{Definition}

\algnewcommand{\algorithmicforeach}{\textbf{for each}}
\algdef{SE}[FOR]{ForEach}{EndForEach}[1]
  {\algorithmicforeach\ #1\ \algorithmicdo}
  {\algorithmicend\ \algorithmicforeach}

\newcommand{\boldheading}[1]{%
    \vspace{0.5em} 
    \noindent\textbf{#1}\hspace{0.1em} 
}

\usepackage{soul} 

\definecolor{vibrantgreen}{RGB}{144, 238, 144} 
\definecolor{vibrantyellow}{RGB}{255, 255, 102}
\definecolor{vibrantblue}{RGB}{202,255, 255}
\definecolor{vibrantred}{RGB}{202,255, 255}

\AtBeginDocument{%
  }

\copyrightyear{2024}
\acmYear{2024}
\setcopyright{acmlicensed}
\acmConference[CIKM '24]{Proceedings of the 33rd ACM International Conference on Information and Knowledge Management}{October 21--25, 2024}{Boise, ID, USA}
\acmBooktitle{Proceedings of the 33rd ACM International Conference on Information and Knowledge Management (CIKM '24), October 21--25, 2024, Boise, ID, USA}
\acmDOI{10.1145/3627673.3679728}
\acmISBN{979-8-4007-0436-9/24/10}
\begin{document}

\title{Calibration-Disentangled Learning and Relevance-Prioritized Reranking for Calibrated Sequential Recommendation}

\author{Hyunsik Jeon}
\affiliation{
	\institution{University of California, San Diego}
	\city{San Diego, CA}
	\country{USA}
}
\email{hyjeon@ucsd.edu}

\author{Se-eun Yoon}
\affiliation{
	\institution{University of California, San Diego}
	\city{San Diego, CA}
	\country{USA}
}
\email{seeuny@ucsd.edu}

\author{Julian McAuley}
\affiliation{
	\institution{University of California, San Diego}
	\city{San Diego, CA}
	\country{USA}
}
\email{jmcauley@ucsd.edu}

\renewcommand{\shortauthors}{Hyunsik Jeon, Se-eun Yoon, and Julian McAuley}

\begin{abstract}
Calibrated recommendation, which aims to maintain personalized proportions of categories within recommendations, is crucial in practical scenarios since it enhances user satisfaction by reflecting diverse interests.
However, achieving calibration in a sequential setting (i.e., calibrated sequential recommendation) is challenging due to the need to adapt to users' evolving preferences.
Previous methods typically leverage reranking algorithms to calibrate recommendations after training a model without considering the effect of calibration and do not effectively tackle the conflict between relevance and calibration during the reranking process.
In this work, we propose \method (\methodfullul), a novel approach for the calibrated sequential recommendation that addresses these challenges.
\method consists of two phases, model training phase and reranking phase.
In the training phase, a backbone model is trained using our proposed calibration-disentangled learning-to-rank loss, which optimizes personalized rankings while integrating calibration considerations.
In the reranking phase, relevant items are prioritized at the top of the list, with items needed for calibration following later to address potential conflicts between relevance and calibration.
Through extensive experiments on four real-world datasets, we show that \method consistently outperforms previous methods in the calibrated sequential recommendation.
Our code is available at \url{https://github.com/jeon185/LeapRec}.
\end{abstract}



\maketitle

\section{Introduction}
\label{sec:intro}

\begin{figure}[t]
	\begin{subfigure}{.485\linewidth}
		\centering
		\includegraphics[width=1\linewidth]{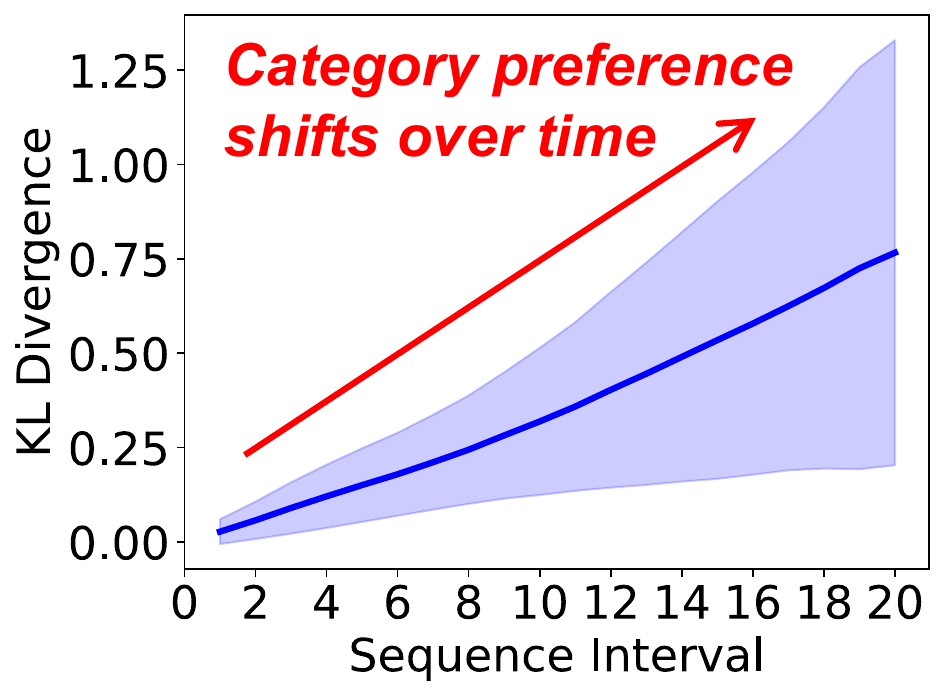}
		\vspace{-1.5em}
		\caption{ML-1M}
		\label{fig:ml-1m_kl}
	\end{subfigure}
    ~
    \begin{subfigure}{.485\linewidth}
        \centering
        \includegraphics[width=1\linewidth]{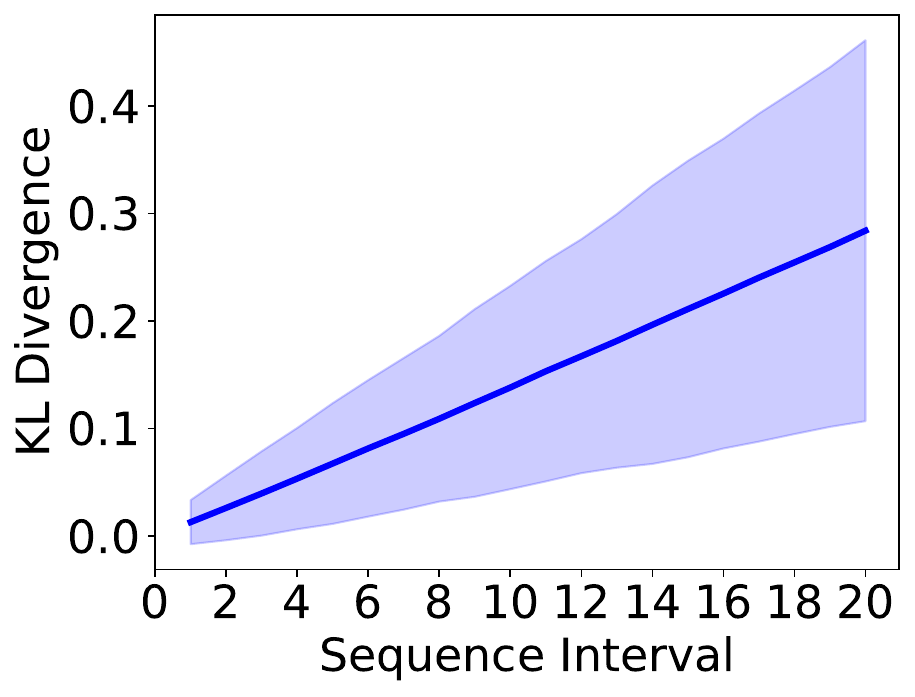}
        \vspace{-1.5em}
        \caption{Steam}
        \label{fig:steam_kl}
    \end{subfigure}
    \caption{
    	KL divergence analysis of user-interacted category distributions over sequence intervals, employing a window size of 20 for each category distribution.
        This plot illustrates the shifts in category preferences over time in real-world datasets: ML-1M and Steam.
    	Detailed data statistics are summarized in Table~\ref{table:datasets}.
    }
    \label{fig:shift}
\end{figure}

Calibrated recommendation aims to reflect a user's diverse interests within a recommendation list by maintaining the proportions of various categories observed in past interactions~\cite{Steck18,SeymenAM21,AbdollahpouriNG23,chen2022dacsr}.
For instance, if a user has historically watched 70\% drama and 30\% action movies, calibrated recommendation should suggest a list of movies maintaining a similar genre ratio.
This problem differs from studies like~\cite{KweonKY22,WeiZHLLWZ22,WangJ23} that define calibration in terms of probability-based user preference estimation, such as estimating how likely a user will prefer an item.
To achieve the calibrated recommendation, two potentially conflicting objectives must be addressed: 1) \emph{relevance}, which aligns with the user's current preferences, and 2) \emph{calibration}, which sustains consistency with their long-term category interests.
This challenge becomes particularly significant in sequential settings (i.e., calibrated sequential recommendation) where users' category preferences shift over time.
These dynamic shifts in preferences are depicted in Figure~\ref{fig:shift}, where the Kullback–Leibler (KL) divergence between category distributions increases as the sequence interval extends, highlighting the intricate challenge of balancing relevance with calibration.

Existing work on calibrated recommendation focuses on post-processing approaches~\cite{Steck18,SeymenAM21,AbdollahpouriNG23}.
Specifically, a recommendation model is first trained to meet the relevance objective; then, the model output is reranked to meet the calibration objective.
The difference between these methods lies in reranking, such as greedy (CaliRec~\cite{Steck18}), mixed integer programming (MIP~\cite{SeymenAM21}), and minimum-cost flow (MCF~\cite{AbdollahpouriNG23}) algorithms.
However, applying calibration during reranking can lead to degradation of accuracy because they do not consider the impacts of calibration during the training phase.
Hence, these methods encounter challenges in maintaining the accuracy of recommendations in the reranking phase, since the changes required for calibration can conflict with the relevance-based ranking criteria, especially in sequential settings where category preference shifts over time as shown in Figure~\ref{fig:shift}.
Recently, DACSR~\cite{chen2022dacsr} introduced an end-to-end model designed to optimize both relevance and calibration during the training phase.
However, DACSR optimizes the calibration across the entire item set to ensure the loss function is differentiable although the primary goal is to calibrate the top-$k$ recommendations.
As a result, DACSR struggles to simultaneously optimize for both accuracy and calibration, particularly for the top-$k$ recommendations.

In this work, we propose \method (\methodfullul), a method that considerably improves upon previous work by novel training and reranking schemes.
First, \method learns to disentangle calibration from relevance during the training phase by optimizing our proposed \textit{calibration-disentangled learning-to-rank} loss.
This approach enables the model to learn rankings adaptable to changes in calibration, allowing for accurate ranking even when calibration is extensively considered.
Consequently, even when calibration adjustments are applied during reranking, the calibration is effectively enhanced without sacrificing much accuracy.
After the training phase, \method further optimizes for relevance and calibration through our proposed \textit{relevance-prioritized reranking}.
This approach prioritizes placing relevant items at the top of the recommendation list while maintaining calibration across the entire list.
This strategy effectively mitigates the risk of excluding items from categories previously unexplored by users, a limitation encountered with previous calibration objectives.

Our contributions are as follows:
\begin{itemize}[leftmargin=6mm]
    \item We propose a novel learning method to effectively disentangle calibration from relevance in the training phase, thus enabling us to facilitate adaptable ranking under varying calibration considerations.
    \item We propose a novel reranking scheme that effectively enhances relevance and calibration by prioritizing relevant items while ensuring consistent calibration across the list.
    \item We conduct extensive experiments on four real-world datasets and show that \method achieves superior performance compared to existing approaches.
\end{itemize}

\section{Preliminaries}
\label{sec:preliminary}
\subsection{Problem Definition}
\label{subsec:problem_definition}
The problem of calibrated sequential recommendation is defined as follows.
Let $\mathcal{U}$, $\mathcal{I}$, and $\mathcal{C}$ be the sets of users, items, and categories, respectively.
Each item $i \in \mathcal{I}$ is associated with a set of categories $\mathcal{C}^i = \{c^i_1, c^i_2, ..., c^i_{N}\}$, where each $c^i_n \in \mathcal{C}$ represents a category of item $i$, and $N$ is the number of categories, varying per item.
For each user $u \in \mathcal{U}$, we have sequential interactions $\mathcal{S}^u = (s^u_1, s^u_2, \dots, s^u_{T})$, where $s^u_t \in \mathcal{I}$ is user $u$'s interacted item at step $t$, and $T$ denotes the length of the sequence which varies among users.
Given sequences $\mathcal{S}^u$ of all users $u \in \mathcal{U}$, our goal is to recommend each user an item list $\mathcal{R}^u = (r_1^u, r_2^u, \dots, r_K^u)$ that is relevant to the user's future needs (i.e., accurate), while reflecting the user's sequential category interests with their appropriate proportions (i.e., sequentially calibrated), at step $T+1$;
$r^u_k \in \mathcal{I}$ is the $k$'th recommended item to user $u$ where lower $k$ indicates higher rank.
The metrics for measuring the degree of calibration are detailed in Section~\ref{subsec:metrics}.

The primary challenge in calibrated sequential recommendation lies in adeptly balancing the often conflicting demands of relevance, which reflects a user's immediate preferences, and calibration, which ensures consistency with their long-term category interests.
This task is further complicated by the constantly evolving nature of user preferences.

\begin{figure}[t]
	\centering
	\includegraphics[width=0.95\linewidth]{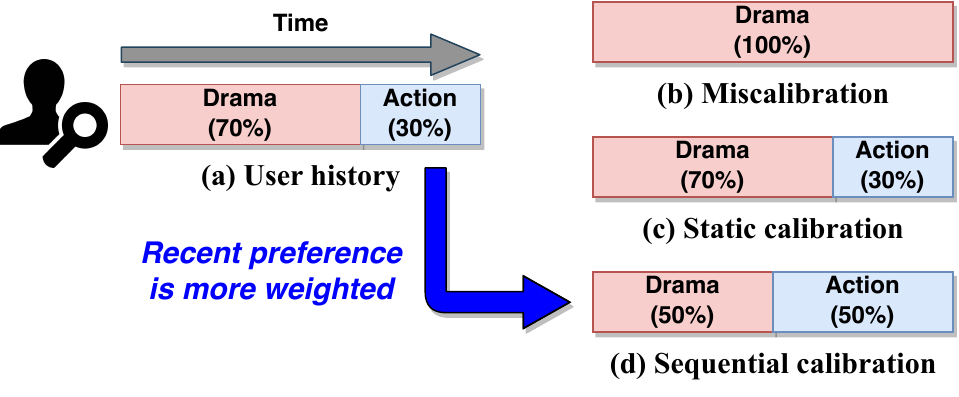}
	\caption{
        Given user history (a), sequential calibration (d) applies more weight to recent category preference, in contrast to recommendation without calibration (b) and static calibration (c).
	}
\label{fig:illustration}
\end{figure}

\subsection{Calibration Metrics for Sequential Recommendation}
\label{subsec:metrics}

The degree of calibration is measured by comparing the category distributions of items in a user's past interactions and items in their recommended list~\cite{Steck18}.
Specifically, it is defined as the divergence between these two distributions (i.e., miscalibration), where a lower value indicates superior calibration performance.
\citet{Steck18} proposed  metrics for miscalibration under various criteria such as whether the user interactions are treated as equally or weighted regarding their recency, as illustrated in Figure~\ref{fig:illustration}.
In this work, we adopt \emph{sequential miscalibration} as our calibration metric, specifically tailored for sequential recommendations.
This metric is described in the following definition.

\begin{definition}[Sequential miscalibration]
\label{df:sequential_miscalibration}
	Given user $u$'s sequential interactions $\mathcal{S}^u$ and the user's recommendation list $\mathcal{R}^u$,
	the sequential miscalibration $\mathcal{S}_{KL}(u)$ is defined as follows:
	\begin{equation}
	\label{eq:kl}
		\mathcal{S}_{KL}(u) = \mathit{KL}(p\|\tilde{q}) = \sum_{c\in\mathcal{C}} p(c|u)\log\frac{p(c|u)}{\tilde{q}(c|u)},
	\end{equation}
	where
	\begin{equation}
	\label{eq:p}
		p(c|u) = \frac{\sum_{s_t^u \in \mathcal{S}^u} \alpha^{T-t} \cdot p(c|s_t^u)}{\sum_{s_t^u \in \mathcal{S}^u} \alpha^{T-t}},
	\end{equation}
	\begin{equation}
	\label{eq:q}
		q(c|u) = \frac{\sum_{r_k^u \in \mathcal{R}^u} p(c|r_k^u)}{|\mathcal{R}^u|},
	\end{equation}
	\begin{equation}
	\label{eq:q_tilde}
		\tilde{q}(c|u) = (1-\beta) q(c|u) + \beta p(c|u),
	\end{equation}
	$\mathit{KL}(\cdot)$ indicates the Kullback–Leibler (KL) divergence between two distributions, $T=|\mathcal{S}^u|$, and $\alpha,\beta\in (0,1)$ are hyperparameters.
\end{definition}

\begin{figure*}[t]
	\centering
	\includegraphics[width=1.0\linewidth]{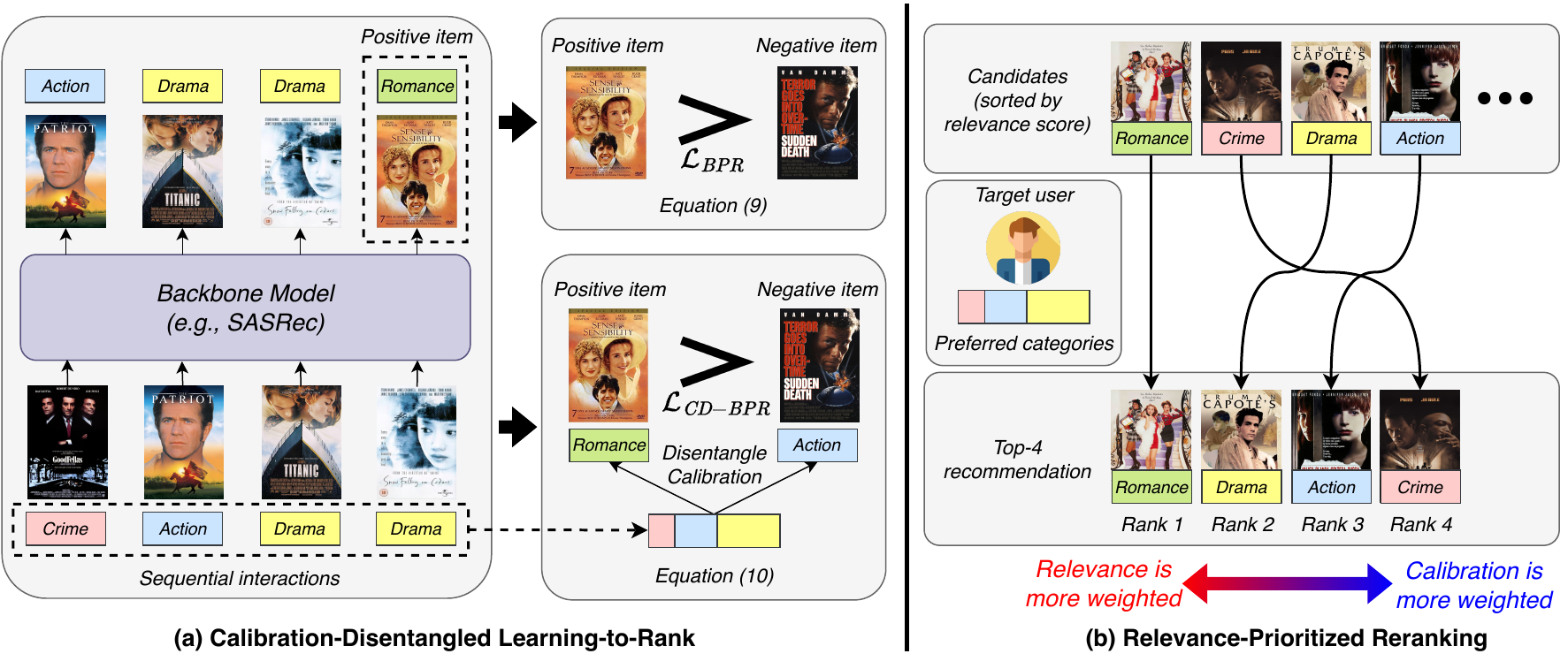}
	\caption{
        \method consists of two phases: 
        (a) \textul{calibration-disentangled learning-to-rank} and (b) \textul{relevance-prioritized reranking}.
        In the first phase, a backbone model is trained to optimize personalized rankings, accommodating both with and without calibration considerations.
        In the second phase, items are greedily added to the recommendation list, where relevance is prioritized at higher ranks and calibration at lower ranks.
	}
\label{fig:overview}
\end{figure*}

In Equations~\eqref{eq:p} and~\eqref{eq:q}, $p(c|s_t^u)$ and $p(c|r_k^u)$ are the category distributions of items $s_t^u,r_k^u \in \mathcal{I}$, respectively.
If an item is associated with multiple categories, each category is equally weighted in the distribution.
In Equation~\eqref{eq:p}, $\alpha$ (e.g., $0.9$) enables us to consider recent interests more weighted to the calibration, thus differentiating this metric from static miscalibration.
In Equation~\eqref{eq:q_tilde}, we modify $q(c|u)$ to $\tilde{q}(c|u) = (1-\beta) q(c|u) + \beta p(c|u)$ with a small value of $\beta$ (e.g., $0.01$), ensuring the KL divergence remains well-defined and does not diverge, as in previous work~\cite{Steck18}.

\section{Proposed Method}
\label{sec:proposed}
In this section, we describe \method (\methodfullul), a novel approach for calibrated sequential recommendation.
Figure~\ref{fig:overview} depicts the overall process of \method, which consists of two phases: model training phase and reranking phase.
During the training phase, \method employs a sequential recommendation model, such as SASRec~\cite{KangM18}, as its backbone.
It optimizes our proposed \emph{calibration-aware learning-to-rank} loss, which is designed to disentangle calibration from relevance, enabling the model to estimate accurate personalized rankings regardless of whether calibration is considered.
During the reranking phase, \method applies our proposed \emph{relevance-prioritized reranking} algorithm.
This algorithm adjusts the model’s output for each user to reduce miscalibration, while prioritizing the most relevant items in the final recommendations.

\subsection{Calibration-Disentangled Learning-to-Rank}
\label{subsec:learning}
The objective of the training phase is to train a sequential model that can predict the probability a user will interact with an item based on their past interactions.
Formally, given a user $u$'s sequence of interactions $\mathcal{S}^u = (s^u_1, s^u_2, \dots, s^u_{T})$ where $s^u_t \in \mathcal{I}$ is the interacted item at step $t$, the model aims to estimate the probability for all items $v\in \mathcal{I}$ at step $T+1$:
\begin{equation}
	p(s^u_{T+1} = v | \mathcal{S}^u).
\end{equation}
Generally, sequential models are trained to increase the gap between the relevance scores of interacted items and non-interacted items using pointwise~\cite{HuKV08}, pairwise~\cite{RendleFGS09} or setwise~\cite{MaoZWDDXH21} losses.
Various frameworks, including Markov Chains~\cite{HeM16a,RendleFS10}, Recurrent Neural Networks (RNN)~\cite{HidasiKBT15,LiRCRLM17,KooJK20,KooJK21}, Convolutional Neural Networks (CNN)~\cite{TangW18,YuanKAJ019}, and self-attention mechanisms~\cite{KangM18,LiWM20,SunLWPLOJ19}, are effectively used in these sequential models, demonstrating high performance in terms of accuracy.

Let $f_\theta(u, i, t)$ represent the relevance score between user $u$ and item $i$ at step $t$, with parameters $\theta$.
Existing calibrated recommendation methods~\cite{Steck18,SeymenAM21,AbdollahpouriNG23} often rely solely on post-processing techniques and thus overlook the potential impact of calibration adjustments on the final ranking order.
However, it is crucial to integrate calibration directly into the training process to anticipate how rankings might change when calibration is applied.
This proactive approach is essential for achieving high performance in both accuracy and calibration.
To illustrate the importance of integrating calibration directly into the training process, consider a scenario where relevance scores indicate a preference for item $i$ over item $j$ based on past interactions (i.e., $f_\theta(u, i, t) > f_\theta(u, j, t)$).
After the training phase, calibration scores may compel the system to rank item $j$ higher than item $i$.
It is difficult to determine whether these items should be reranked, since the degree to which item $i$ is more relevant than item $j$ becomes uncertain when calibration is taken into account.
Thus, ensuring that the model can maintain consistent rankings even after calibration adjustments during reranking is crucial for the calibrated sequential recommendation.

To address this issue, we propose a calibration-disentangled learning-to-rank, a model-agnostic learning approach.
For brevity, let $f_\theta(u,i,t) := r_{i,t}^u$.
Suppose user $u$ interacted with item $i$ instead of item $j$ at step $t$.
This interaction indicates that the user prefers item $i$ over item $j$, even with category preference taken into account. Thus, at the recommendation step $t$, it is crucial to recommend item $i$ to user $u$ over item $j$, even subsequent to calibration.
Existing methods learn a preference order without considering the category preference as follows: 
\begin{equation}
\label{eq:bpr_ranking}
    r_{i,t}^u > r_{j,t}^u.
\end{equation}
This order does not guarantee that item $i$ will be prioritized over item $j$ after considering calibration.
Hence, we propose to disentangle calibration from pairwise ranking, which learns that item $i$ is preferred over item $j$ even with calibration considerations:
\begin{equation}
\label{eq:cabpr_ranking}
    r_{i,t}^u - \mathit{KL}(p_{t-1}^u || p_{i}) > r_{j,t}^u - \mathit{KL}(p_{t-1}^u || p_{j}),
\end{equation}
where $p_{t-1}^u$ is the sequential category preference of user $u$ at step $t-1$, and $p_i$ and $p_j$ are the category distributions of items $i$ and $j$, respectively.
$\mathit{KL}(\cdot)$ is the miscalibration score computed by Equation~\eqref{eq:kl}.
By integrating calibration directly into the training phase, our method trains the model to optimize relevance while being aware of calibration needs.
This ensures that adjustments made for calibration in the reranking phase do not negatively impact the relevance of the recommendations, thereby maintaining accuracy when enhancing calibration.
To learn these two pairwise ranking losses (Equations~\eqref{eq:bpr_ranking} and~\eqref{eq:cabpr_ranking}), we propose the \emph{calibration-disentangled learning-to-rank} loss, which extends the Bayesian Personalized Ranking (BPR) loss as follows:
\begin{equation}
\label{eq:loss}
	\mathcal{L} = \mathcal{L}_{\mathit{BPR}} + \gamma \mathcal{L}_{\mathit{CD-BPR}},
\end{equation}
where
\begin{equation}
\label{eq:bpr}
	\mathcal{L}_{\mathit{BPR}} = \sum_{u\in\mathcal{U}} \sum_{t\in[1,\dots,T]}-\log\sigma\left(r_{i,t}^u - r_{j,t}^u\right),
\end{equation}
\begin{equation}
\label{eq:cabpr}
\begin{split}
	&\mathcal{L}_{\mathit{CD-BPR}} =\\
    &\sum_{u\in\mathcal{U}} \sum_{t\in[1,\dots,T]} -\log\sigma\left(r_{i,t}^u - \mathit{KL}(p_{t-1}^u || p_{i}) - r_{j,t}^u + \mathit{KL}(p_{t-1}^u || p_{j})\right),
\end{split}
\end{equation}
where $\sigma(\cdot)$ is the sigmoid function and $\gamma \in \mathbb{R}$ is a hyperparameter that determines the importance of $\mathcal{L}_{\mathit{CD-BPR}}$.
The calibration-disentangled learning-to-rank does not require learning additional parameters, thus avoiding an increase in model complexity.

\begin{figure}[t]
	\centering
	\includegraphics[width=0.95\linewidth]{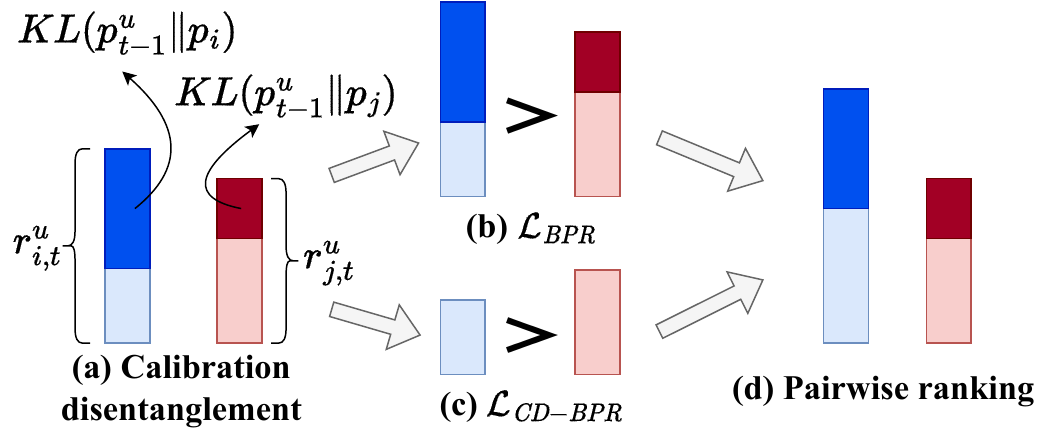}
	\caption{
        Illustrative example of calibration-disentangled learning-to-rank. Initially, miscalibration scores are disentangled from relevance scores (a). Then, we train the model based on both $\mathcal{L}_{\mathit{BPR}}$ (b) and $\mathcal{L}_{\mathit{CD-BPR}}$ (c), resulting in a personalized pairwise ranking considering calibration (d).
	}
\label{fig:cd-ltr}
\end{figure}

The key aspect of calibration-disentangled learning-to-rank is its ability to dynamically adjust relevance scores by incorporating calibration scores based on the user's category preferences.
As illustrated in Figure~\ref{fig:cd-ltr}, this method initially separates the miscalibration scores from relevance scores (Figure~\ref{fig:cd-ltr} (a)).
For instance, if the categories of item $i$ significantly diverge from user $u$'s preferences before $t$, the model $f_\theta$ is prompted to increase the relevance score $r_{i,t}^u$ (Figure~\ref{fig:cd-ltr} (d)) to compensate for the high miscalibration score $\mathit{KL}(p_{t-1}^u | p_i)$, using $\mathcal{L}_{\mathit{CD-BPR}}$ (Figure~\ref{fig:cd-ltr} (c)).
This approach differs from previous methods that rely solely on $\mathcal{L}_{\mathit{BPR}}$, where personalized ranking might change unpredictably after calibration during reranking.
However, relying only on $\mathcal{L}_{\mathit{CD-BPR}}$ also can lead to issues in personalized rankings.
For example, consider a case where $r_{i,t}^u - \mathit{KL}(p_{t-1}^u || p_{i})$ is larger than $r_{j,t}^u - \mathit{KL}(p_{t-1}^u || p_{j})$, but $r_{i,t}^u$ is lower than $r_{j,t}^u$.
In such case, $\mathcal{L}_{\mathit{CD-BPR}}$ prioritizes item $i$ over item $j$ when calibration is considered, but it may overlook genuine relevance if calibration is not considered.
This case verifies the necessity of integrating both $\mathcal{L}_{\mathit{BPR}}$ and $\mathcal{L}_{\mathit{CD-BPR}}$.

\renewcommand{\algorithmicrequire}{\textbf{Input:}}
\renewcommand{\algorithmicensure}{\textbf{Output:}}
\begin{algorithm}[t]
    \caption{Overall process of \method}
    \label{al:overall}
    \begin{algorithmic}[1]
        \Require $\{\mathcal{S}^u: u\in\mathcal{U}\}$ and hyperparameters ($\alpha$, $\beta$, $\gamma$, and $\lambda$)
        \Ensure $\{\mathcal{R}^u: u\in\mathcal{U}\}$
            \While{stop condition is not met} \Comment{Train a backbone}
                \For{$u\in \mathcal{U}$}
                    \State Optimize $f_\theta$ to minimize $\mathcal{L}$ \Comment{Equation~\eqref{eq:loss}}
                \EndFor
            \EndWhile

            \For{$u\in \mathcal{U}$} \Comment{Rerank}
                \State{$\mathcal{C}^u = \{\}$}
                \For{$k \in [1,K]$}
                    \State $i \leftarrow \max_{i \in \mathcal{I}\setminus \mathcal{C}^u} (1-\lambda^{1/k})r^u_{i,T+1} - \lambda^{1/k} \Delta \mathcal{S}_{KL} (i | \mathcal{C}^u)$
                    \State $\mathcal{C}^u \leftarrow \mathcal{C}^u \cup \{i\}$
                \EndFor
                \State $\mathcal{R}^u \leftarrow \mathcal{C}^u$
            \EndFor
    \end{algorithmic}
\end{algorithm}

\begin{savenotes}
	\begin{table*}
		\centering
		\caption{Summary of four real-world datasets used in this work.}
		\label{table:datasets}
		\begin{threeparttable}
		\begin{tabular}{l|rrrrrrr}
			\toprule
			\textbf{Dataset} & \textbf{\# Users} & \textbf{\# Items} & \textbf{\# Categories} & \textbf{\# Interactions} & \textbf{Avg. sequence len.} & \textbf{User-item density} & \textbf{Avg. \# categories}\\
			\midrule
			ML-1M & 6,038 & 3,883 & 18 & 575,281 & 95.28 & 0.0245 & 1.6503\\
			Goodreads & 16,765 & 25,474 & 10 & 954,958 & 56.96 & 0.0022 & 3.6269\\
            Grocery & 54,882 & 39,853 & 26 & 438,681 & 7.99 & 0.0002 & 1.0000\\
            Steam & 242,223 & 14,419 & 22 & 2,732,749 & 11.29 & 0.0008 & 2.6242\\
			\bottomrule
		\end{tabular}
        \end{threeparttable}
    \end{table*}
\end{savenotes}

\subsection{Relevance-Prioritized Reranking}
\label{subsec:reranking}
The reranking phase aims to maximize both accuracy and calibration using the trained model $f_\theta$. 
The goal is to recommend user $u$ a list of items $\mathcal{R}^u = (r_1^u, r_2^u, \dots, r_K^u)$ where $r_k^u \in \mathcal{I}$ is the $k$'th recommended item.
Reranking generates the list by selecting the most suitable items among the candidates.
The main challenge in the reranking phase is to measure which item is the best for the user at each step, considering both accuracy and calibration.

From Figure~\ref{fig:shift}, we observe that a user is likely to interact with an item that is associated with a category the user has not preferred before.
For instance, a user who predominantly watches action movies might develop an interest in romance movies due to temporal factors.
In this case, it is necessary to ensure that items that users like are recommended regardless of their category to satisfy the user's future needs.
However, naively using weighted sum~\cite{Steck18,SeymenAM21,AbdollahpouriNG23} may not adequately handle such cases.
In the example, the romance movie may have a low overall score despite its high relevance due to the high miscalibration score, since it contrasts with the user's past category preferences.
This conflict of relevance and calibration should be treated as a critical issue in the calibrated sequential recommendation, yet it has not been thoroughly addressed in previous works~\cite{Steck18,SeymenAM21,AbdollahpouriNG23,chen2022dacsr}.

To address this challenge, our reranking strategy prioritizes a user's emerging interests by integrating both relevance and calibration but favoring relevance in the higher ranks of the recommendation list.
If we consider that the backbone model $f_\theta$ is trained to predict the user's evolving preferences based on a sequential model, we can infer that the relevance scores from the model are more closely related to a user's emerging interests.
Thus, we propose the \emph{relevance priority} property for the reranking algorithm as follows:
\begin{property}[Relevance priority]
\label{prop:relevance_priority}
	In higher-ranked recommendations, relevance should be prioritized over calibration.
\end{property}
\noindent Reranking based on this property prevents potentially relevant items from being lower ranked (or excluded) due to calibration constraints, thereby offering a more accurate reflection of the user's evolving interests.
Such prioritized approach has recently been explored in previous work on multi-objective recommendation, demonstrating its effectiveness~\cite{JeonKLLK23}.

To apply Property~\ref{prop:relevance_priority} in the reranking algorithm, we propose a simple yet effective objective function for each user $u$ as follows:
\begin{equation}
\label{eq:objective}
	\max_{\mathcal{R}^u, |\mathcal{R}^u|=K}\left((1 - \lambda^{1/k}) \sum_{i \in \mathcal{R}^u} r_{i,T+1}^u - \lambda^{1/k} \mathcal{S}_{\mathit{KL}}(u)\right),
\end{equation}
where $\lambda \in [0,1]$ is a balancing hyperparameter between relevance and calibration,
$k\in[1,K]$ indicates the position in the recommendation list, 
$r_{i,T+1}^u$ is the relevance score of item $i$ for user $u$ at step $T+1$ (i.e., $f_\theta(u,i,T+1)$),
and $\mathcal{S}_{\mathit{KL}}(u)$ is sequential miscalibration which is defined in Equation~\eqref{eq:kl}.
Smaller $k$ (i.e., higher ranking) assigns more weight to relevance and larger $k$ (i.e., lower ranking) to calibration.
Hence, the objective function satisfies the relevance priority property.
Moreover, we leverage sequential miscalibration $\mathcal{S}_{\mathit{KL}}(u)$ rather than static miscalibration used in most previous methods~\cite{SeymenAM21,AbdollahpouriNG23,chen2022dacsr} to consider the recent category preferences in measuring the degree of calibration.

Finding the optimal recommendation list from Equation~\eqref{eq:objective} is a combinatorial optimization problem and NP-hard.
Thus, we adopt a greedy approach, which is fast and effective, to optimize the objective function.
Specifically, at the $k$'th recommendation for user $u$, we select an item that maximizes the gain of scores among items that are yet to be selected as follows:
\begin{equation}
\label{eq:greedy}
    \max_{i \in \mathcal{I}\setminus \mathcal{C}^u} \left( (1-\lambda^{1/k})r^u_{i,T+1} - \lambda^{1/k} \Delta \mathcal{S}_{\mathit{KL}} (i | \mathcal{C}^u) \right),
\end{equation}
where $\mathcal{C}^u \subseteq \mathcal{R}^u$ is the current recommendation list for user $u$,
and $\Delta \mathcal{S}_{\mathit{KL}} (i | \mathcal{C}^u)$ indicates the difference of $\mathcal{S}_{\mathit{KL}}(u)$ when item $i$ is added to the current recommendation list $\mathcal{C}^u$.

\subsection{Overall Process of \method}
\label{subsec:process}
The overall process of \method consists of the backbone model training phase and the reranking phase.
Algorithm~\ref{al:overall} shows how \method trains the backbone model and reranks the results.
Given users' sequential interactions $\{\mathcal{S}^u: u\in\mathcal{U}\}$, \method returns recommendation lists $\{\mathcal{R}^u: u\in\mathcal{U}\}$ for all users.
In lines 1 to 5, \method trains a backbone model by minimizing the loss in Equation~\eqref{eq:loss} for predefined epochs (e.g., 100);
in line 2, we adopt a mini-batch training in our practical implementation.
Then, in lines 6 to 13, \method greedily selects $K$ items for each user considering both relevance and sequential miscalibration.

\section{Experiments}
\label{sec:experiments}
\begin{figure*}[t]
        \centering
        \begin{subfigure}{0.9\linewidth}
                \centering
                \includegraphics[width=1\linewidth]{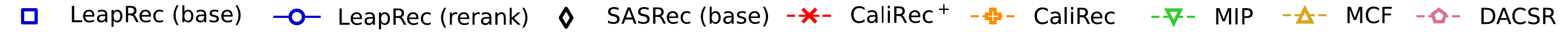}
                \label{fig:overall_legend}
                \vspace{-1.5em}
        \end{subfigure}
        \\
        \begin{subfigure}{.25\linewidth}
                \centering
                \includegraphics[width=1\linewidth]{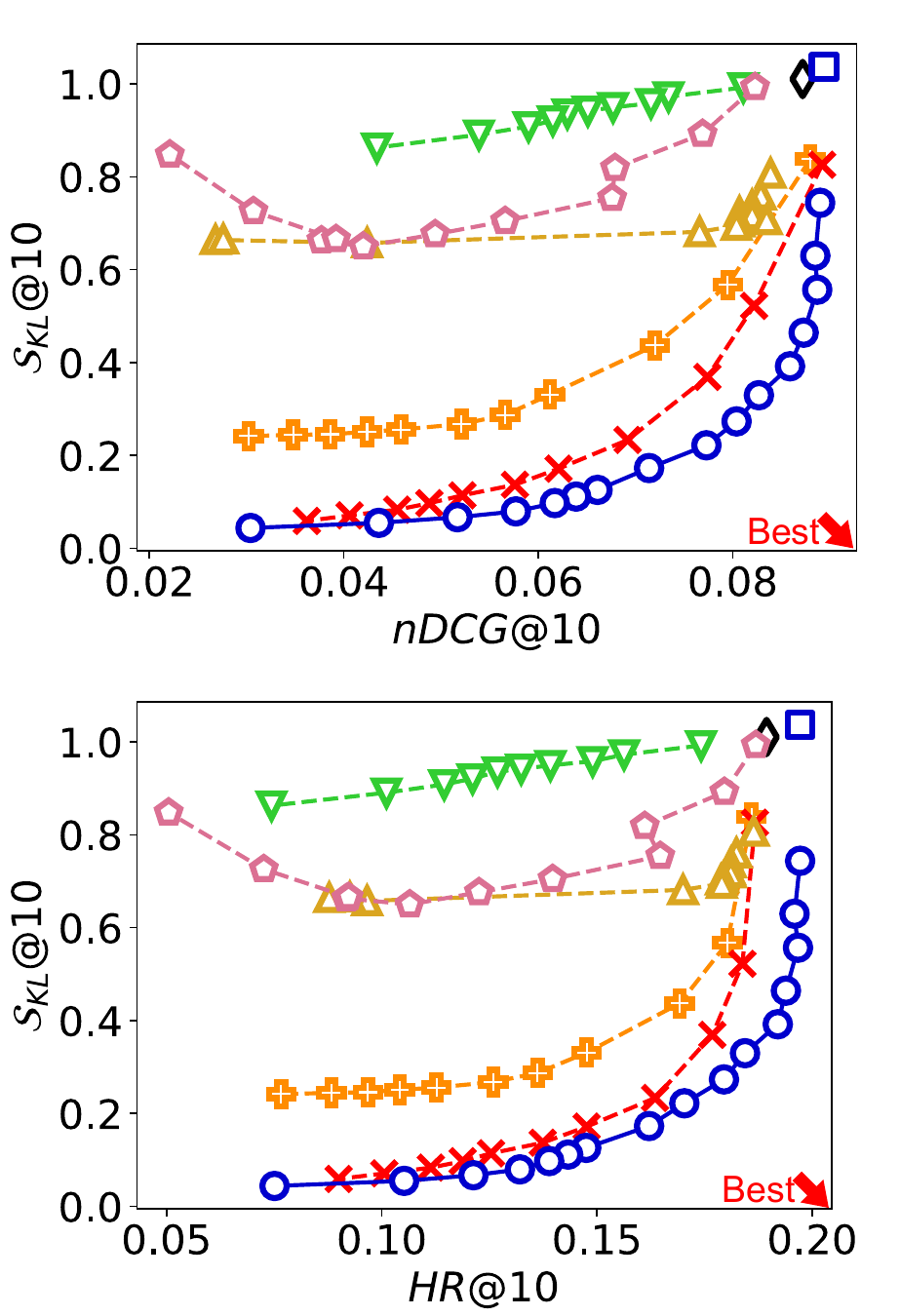}
                \vspace{-2em}
                \caption{ML-1M}
                \label{fig:overall_ml-1m}
        \end{subfigure}
        ~
        \begin{subfigure}{.25\linewidth}
                \centering
                \includegraphics[width=1\linewidth]{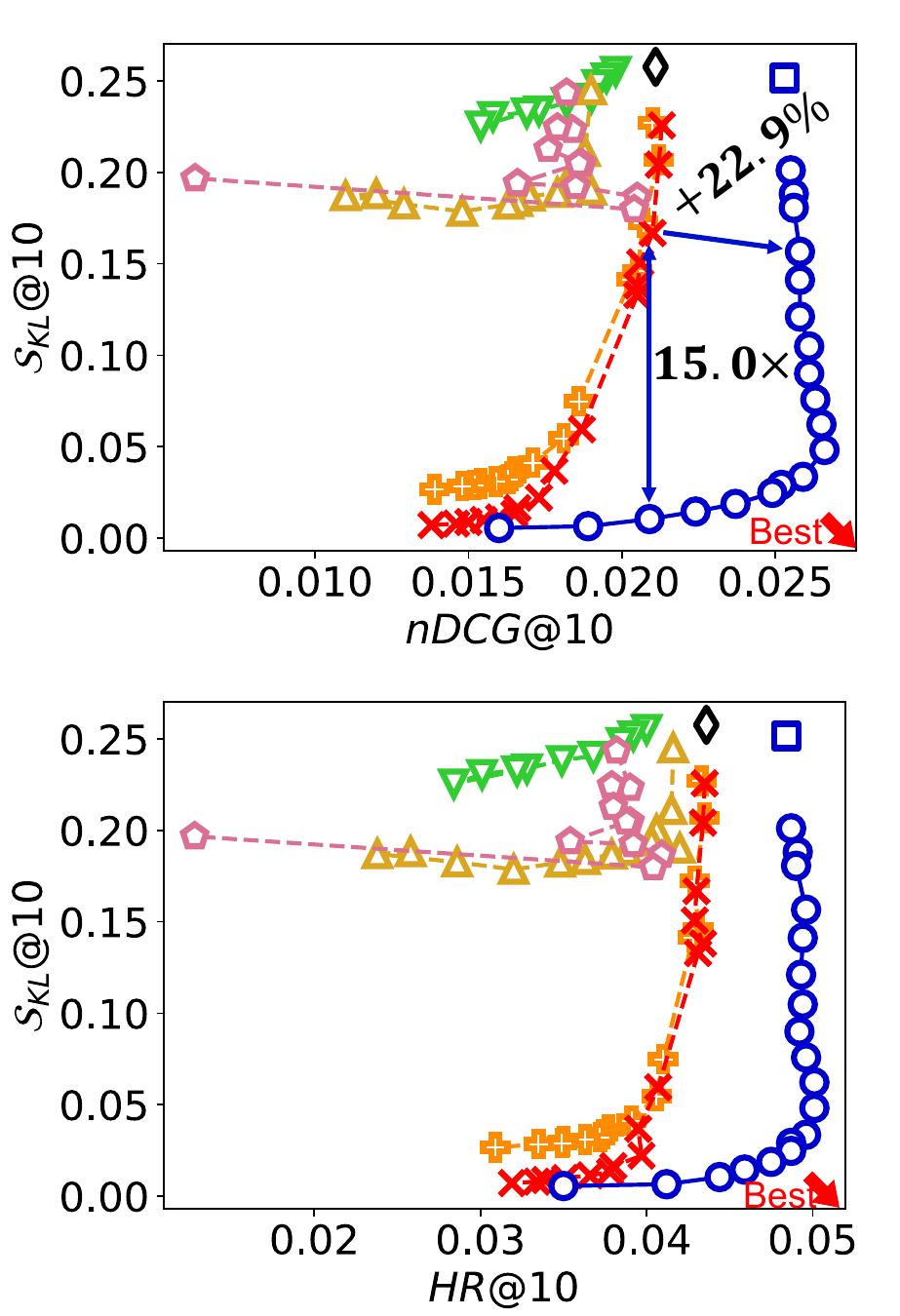}
                \vspace{-2em}
                \caption{Goodreads}
                \label{fig:overall_goodreads}
        \end{subfigure}
        ~
        \begin{subfigure}{.25\linewidth}
                \centering
                \includegraphics[width=1\linewidth]{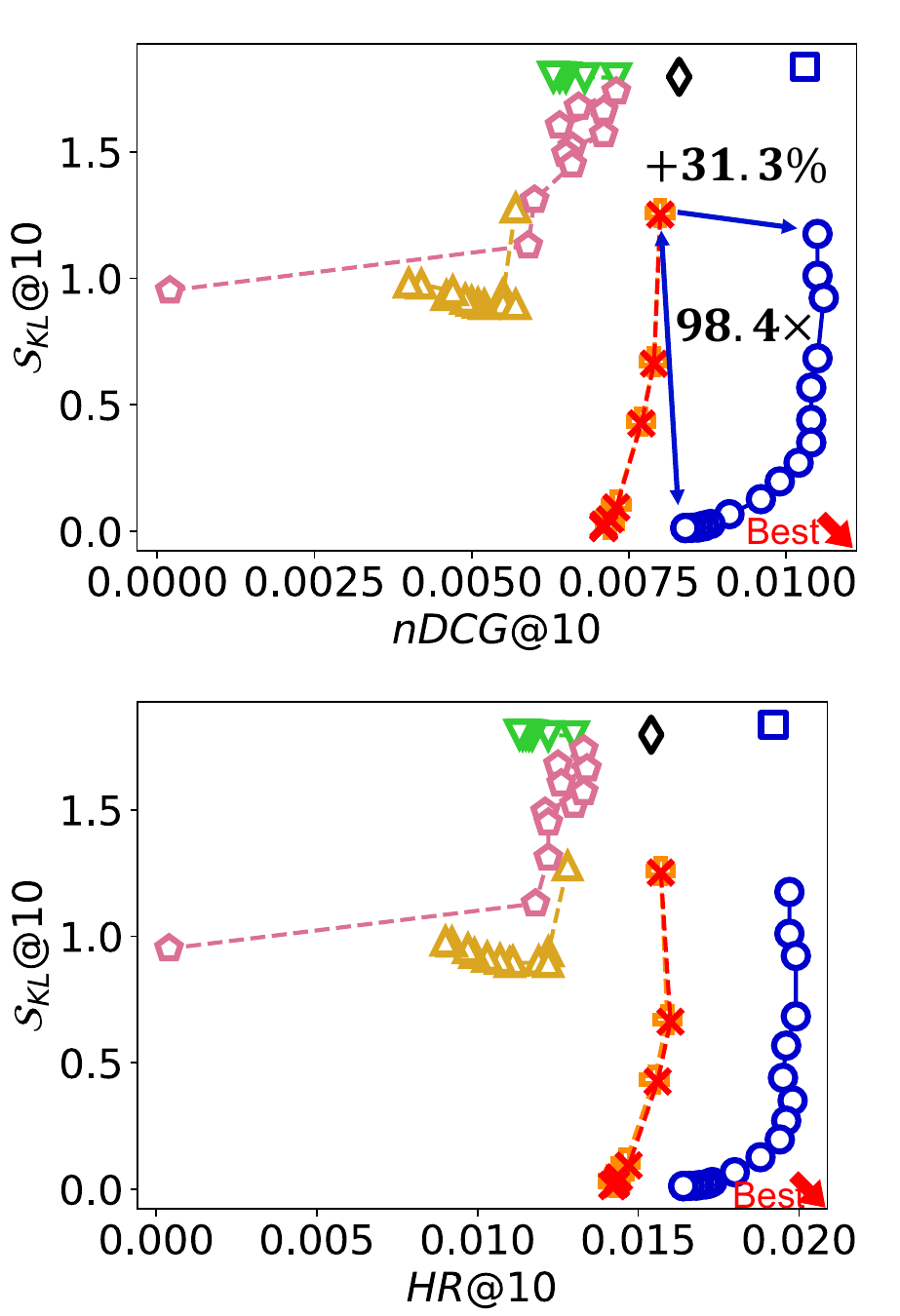}
                \vspace{-2em}
                \caption{Grocery}
                \label{fig:overall_grocery}
        \end{subfigure}
        ~
        \begin{subfigure}{.25\linewidth}
                \centering
                \includegraphics[width=1\linewidth]{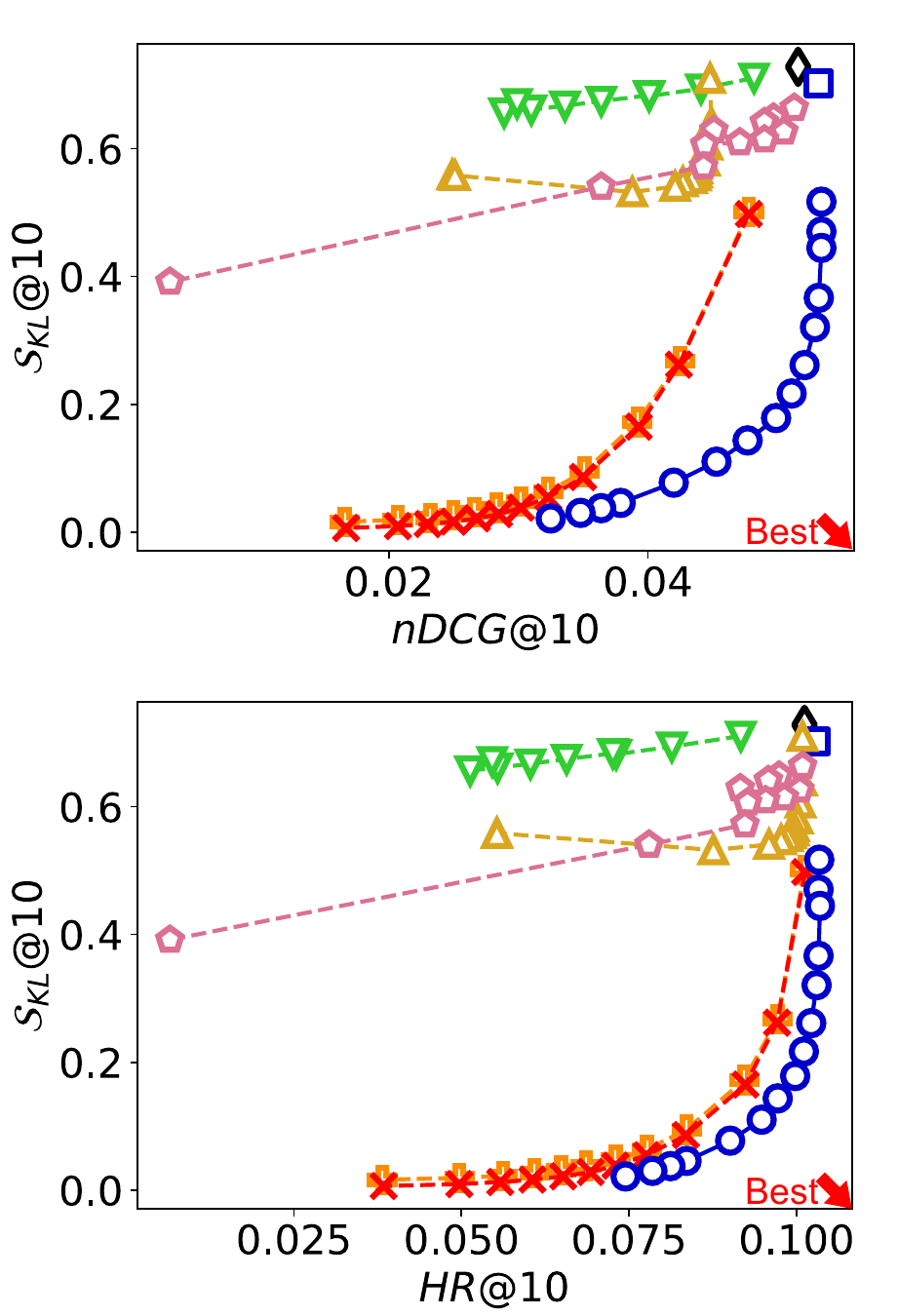}
                \vspace{-2em}
                \caption{Steam}
                \label{fig:overall_steam}
        \end{subfigure}
        \caption{
        	Trade-off comparison of \method and baselines.
        	The first row shows $nDCG@10$ vs. $\mathcal{S}_{KL}@10$, and the second row shows $HR@10$ vs. $\mathcal{S}_{KL}@10$.
            \method outperforms the baselines on four real-world datasets, drawing better trade-off curves between accuracy and calibration.
        }
        \label{fig:overall}
\end{figure*}

In this section, we conduct experiments to answer the following questions.

\begin{itemize}[leftmargin=6mm]
	\item[Q1.] \textbf{Performance comparison (Section~\ref{subsec:performance_comparison}).}
        Does \method provide better trade-off between accuracy and calibration compared to competitors?
        How efficient and fast is \method compared to competitors in generating recommendations?
	\item[Q2.] \textbf{Ablation study (Section~\ref{subsec:ablation_study}).}
        Do the main components in \method help improve the performance?
    \item[Q3.] \textbf{Effect of the balancing hyperparameter (Section~\ref{subsec:equity}).}
        How does the balancing hyperparameter $\lambda$, which is the key factor in enhancing calibration, affect the overall recommendation?
	\item[Q4.] \textbf{Case study (Section~\ref{subsec:qaulitative_analysis}).}
        How \method recommend a list of items considering both relevance and calibration?
\end{itemize}

\subsection{Experimental Setting}
\label{subsec:experimental_settings}

\subsubsection{Datasets}
We evaluate \method and other methods using four real-world datasets from distinct domains: movies (ML-1M\footnote{\url{https://grouplens.org/datasets/movielens/1m}}~\cite{HarperK16}), books (Goodreads\footnote{\url{https://cseweb.ucsd.edu/~jmcauley/datasets.html\#goodreads}}~\cite{WanM18}), grocery products (Grocery\footnote{\url{https://cseweb.ucsd.edu/~jmcauley/datasets.html\#amazon_reviews}}~\cite{HeM16b,McAuleyTSH15}), and video games (Steam\footnote{\url{https://cseweb.ucsd.edu/~jmcauley/datasets.html\#steam_data}}~\cite{PathakGM17}).
These datasets are chosen for their diversity in domain, sparsity level, and number of categories, as detailed in Table~\ref{table:datasets}.
In ML-1M dataset, we follow previous work~\cite{Steck18,AbdollahpouriNG23} and consider only ratings of four stars and above, simulating positive feedback.

\subsubsection{Backbone Model}
Unless otherwise stated, we consider SASRec~\cite{KangM18} as a backbone framework which has shown its superior performance compared with other frameworks in comprehensive experiments~\cite{KangM18,KlenitskiyV23}.
However, the calibration-disentangled learning-to-rank approach is open to other sequential recommendation frameworks such as GRU4Rec~\cite{HidasiKBT15}, Caser~\cite{TangW18}, and BERT4Rec~\cite{SunLWPLOJ19} since it is a model-agnostic learning approach.

\subsubsection{Baseline Methods}
We compare \method with the following four existing calibration recommendation methods.
\begin{itemize}[leftmargin=*]
    \item \textbf{CaliRec}~\cite{Steck18} is a post-processing approach that reranks the output of a backbone model using a greedy algorithm, optimizing for static calibration.
    \item \textbf{CaliRec$^+$}~\cite{Steck18} differs from CaliRec in reranking algorithm where it optimizes for sequential calibration instead of static calibration. We adopt $\mathcal{S}_{\mathit{KL}}(u)$ as in \method for the miscalibration score.
    \item \textbf{MIP}~\cite{SeymenAM21} is a post-processing approach that utilizes mixed integer programming, focusing on achieving static calibration.
    \item \textbf{MCF}~\cite{AbdollahpouriNG23} is a post-processiong approach that employs a minimum-cost flow algorithm in its reranking phase to adjust for static calibration.
    \item \textbf{DACSR}~\cite{chen2022dacsr} is an end-to-end approach that simultaneously targets accuracy and static calibration optimization.
\end{itemize}

Each method uses hyperparameter $\lambda$ to balance between accuracy and calibration.

\begin{table}[t]
    \centering
    \small
    \caption{Comparative analysis of computational complexity for reranking algorithms between \method and baselines. Variables include $n$ (number of users), $m$ (number of items), $k$ (recommendation list size), and $c$ (number of categories).}
    \label{table:complexity}
    \begin{tabular}{c|ccc}
        \toprule
        \multirow{3}{*}{\textbf{Method}} & \textbf{\method (ours)} & \multirow{3}{*}{\textbf{MIP}} & \multirow{3}{*}{\textbf{MCF}}\\
        & \textbf{CaliRec} & &\\
        & \textbf{CaliRec$^+$} & &\\
        \midrule
        Complexity & $\mathcal{O}(nkm)$ & $\mathcal{O}( n m^k)$ & $\mathcal{O}( n (k+c) m^2 \log m)$\\
        \bottomrule
    \end{tabular}
\end{table}

\vspace{-1em}
\subsubsection{Experimental Process}
We follow the \emph{leave-one-out} protocol as established by prior studies~\cite{WangHWCSO19,SunLWPLOJ19,abs-2210-15460,JeonKLLK23,KimJLK23}.
For each user $u$, we split their historical interaction sequence $\mathcal{S}^u$ into three parts: the most recent interaction for testing, the second most recent one for validation, and all earlier interactions for training.
For \method and the baseline methods (CaliRec, MIP, MCF, and DACSR), we conduct training of models for $200$ epochs on ML-1M dataset ($100$ epochs on the other datasets) and choose the models with the best validation performance.
Subsequently, we apply each method's specific reranking strategy (except for DACSR), adjusting the balancing hyperparameter $\lambda$ to draw trade-off curves between relevance and calibration.

\vspace{-1em}
\subsubsection{Evaluation Metrics}
We evaluate the performance in two criteria accuracy and calibration, by investigating the trade-off curve between them.
We use hit ratio (HR$@K$) and normalized discounted cumulative gain (nDCG$@K$) metrics to measure the accuracy.
Given top-$K$ recommendation lists, HR$@K$ measures whether the lists contain the ground-truth items, and nDCG$@K$ weighs the rank of ground-truth items in the list.
We use sequential miscalibration ($S_{KL}@K$) metrics to measure the calibration (see Section~\ref{subsec:metrics} for details).
Higher HR$@K$ and nDCG$@K$ indicate better performance, while lower $S_{KL}@K$ indicates better performance.
We set $K$ to $10$ in our experiments.

\begin{table}[t]
    \setlength\tabcolsep{2.5pt}
    \centering
    \small
    \caption{Comparison of running times (in seconds) for reranking algorithms between \method and baselines across various datasets. Bold indicates the fastest record in each row.}
    \label{table:time}
    \begin{tabular}{l|ccccc}
        \toprule
        \textbf{Method} & \textbf{\method (ours)} & \textbf{CaliRec} & \textbf{CaliRec$^+$} & \textbf{MIP} & \textbf{MCF}\\
        \midrule
        ML-1M & \textbf{24} & \textbf{24} & \textbf{24} & 1,793 & 330\\
        Goodreads & 232 & \textbf{228} & 229 & 14,018 & 1,076\\
        Grocery & 2,808 & 2,816 & \textbf{2,799} & 18,176 & 3,015\\
        Steam & \textbf{3,925} & 4,121 & 3,944 & 92,245 & 16,256\\
        \bottomrule
    \end{tabular}
\end{table}

\begin{figure*}[t]
        \centering
        \begin{subfigure}{0.5\linewidth}
                \centering
                \includegraphics[width=1\linewidth]{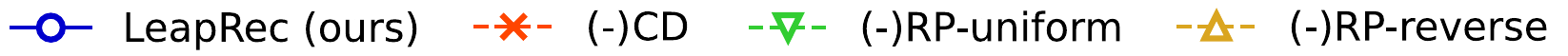}
                \label{fig:ablation_legend}
                \vspace{-1.5em}
        \end{subfigure}
        \\
        \begin{subfigure}{.25\linewidth}
                \centering
                \includegraphics[width=1\linewidth]{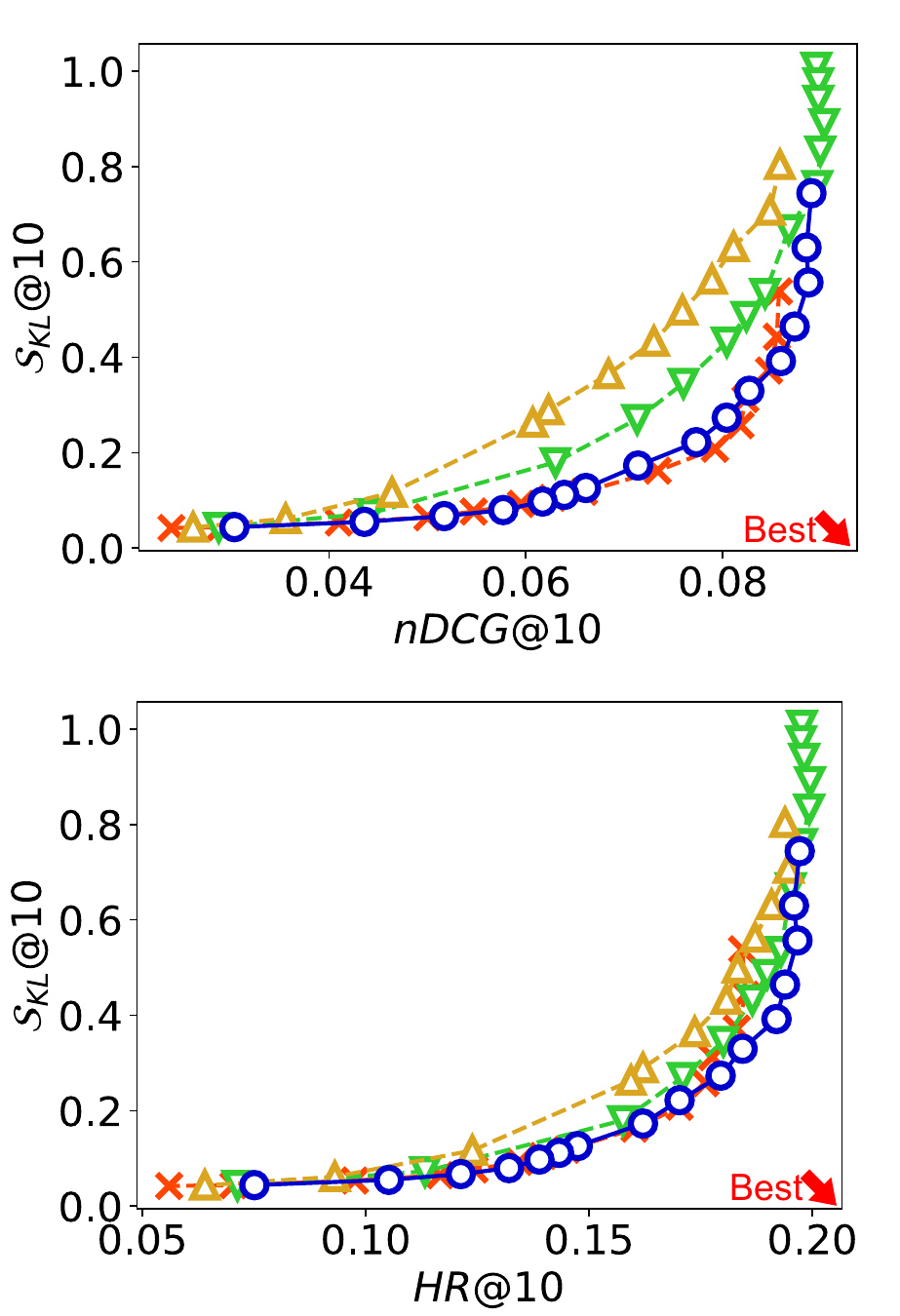}
                \vspace{-2em}
                \caption{ML-1M}
                \label{fig:ablation_ml-1m}
        \end{subfigure}
        ~
        \begin{subfigure}{.25\linewidth}
                \centering
                \includegraphics[width=1\linewidth]{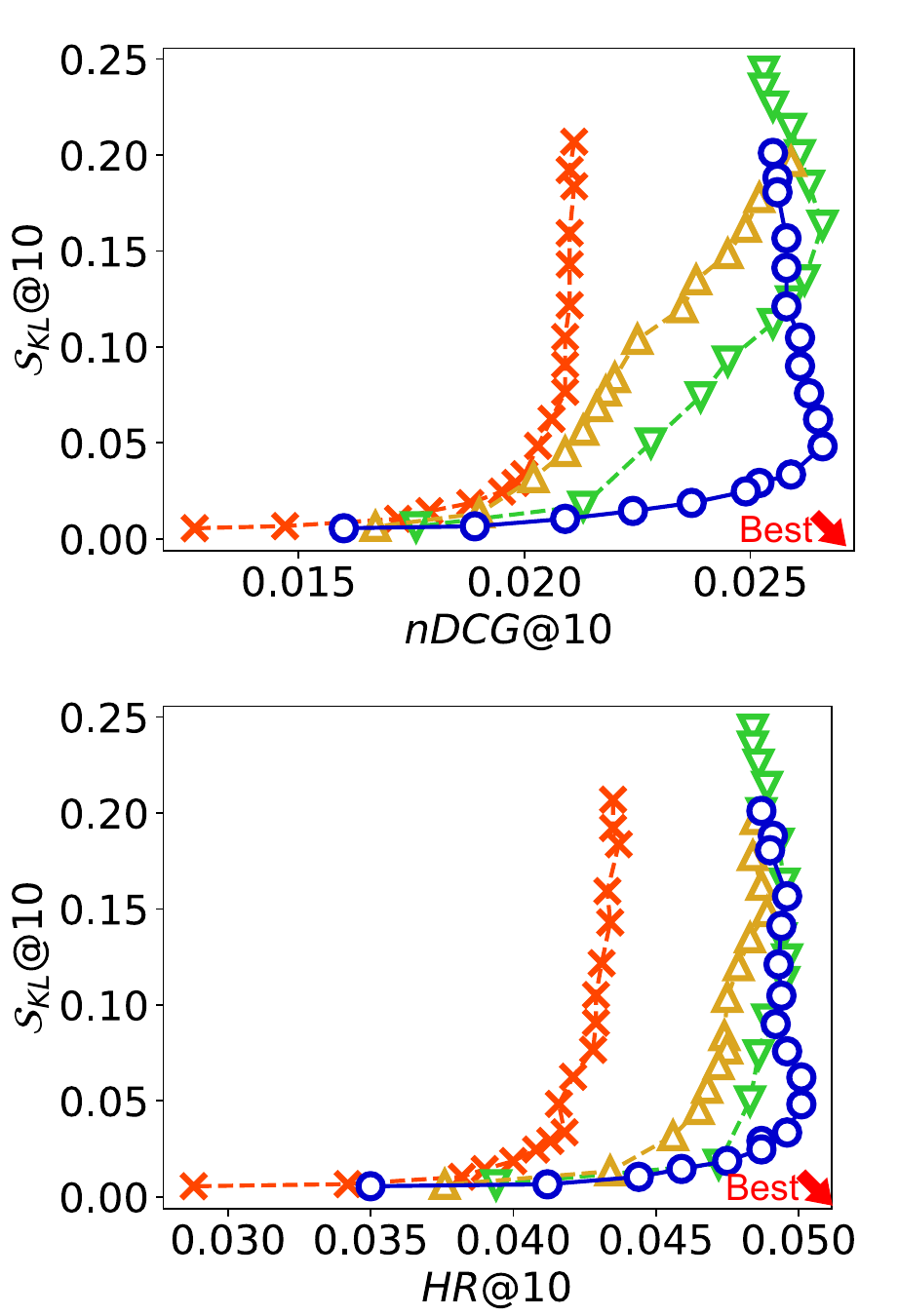}
                \vspace{-2em}
                \caption{Goodreads}
                \label{fig:ablation_goodreads}
        \end{subfigure}
        ~
        \begin{subfigure}{.25\linewidth}
                \centering
                \includegraphics[width=1\linewidth]{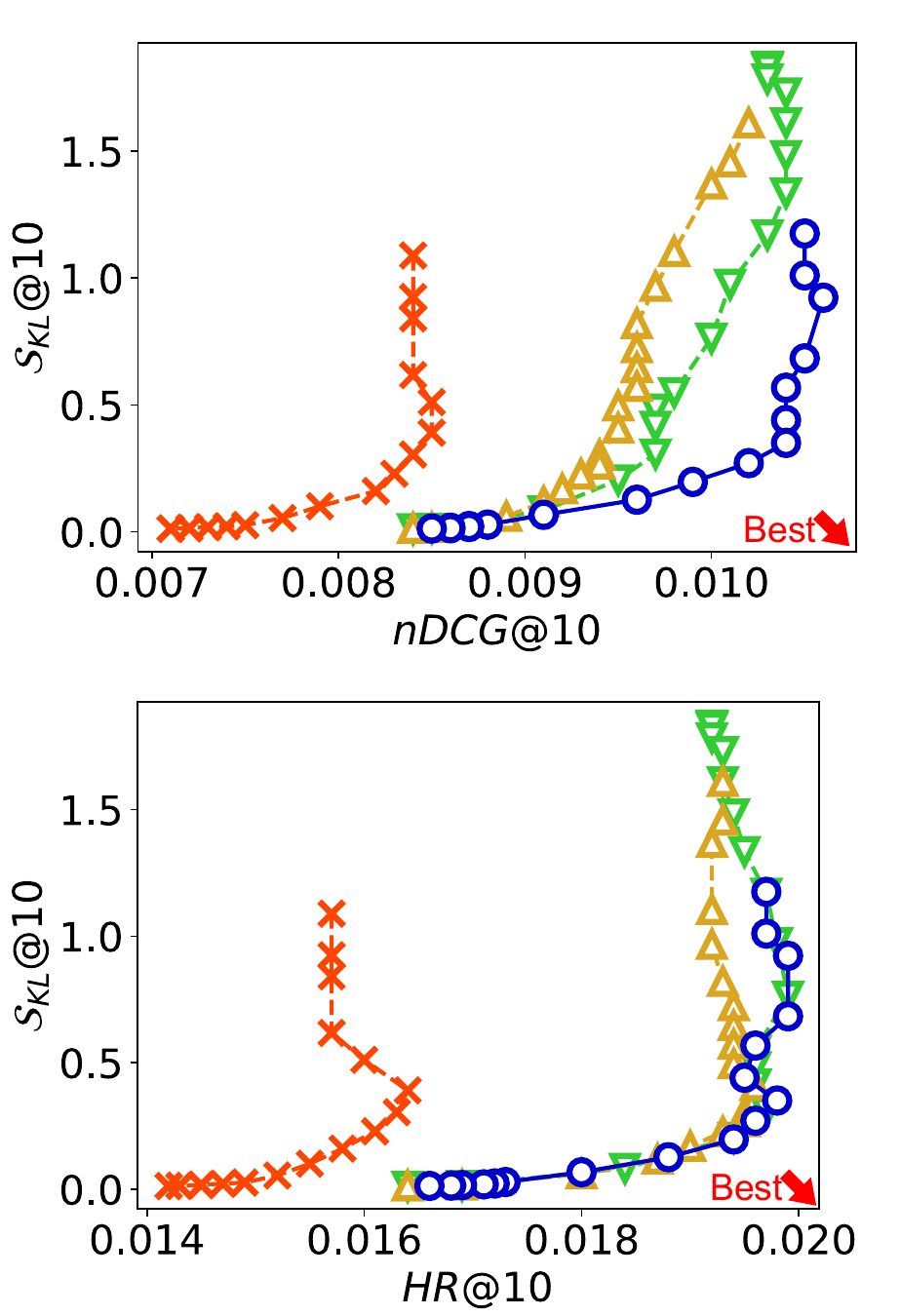}
                \vspace{-2em}
                \caption{Grocery}
                \label{fig:ablation_grocery}
        \end{subfigure}
        ~
        \begin{subfigure}{.25\linewidth}
                \centering
                \includegraphics[width=1\linewidth]{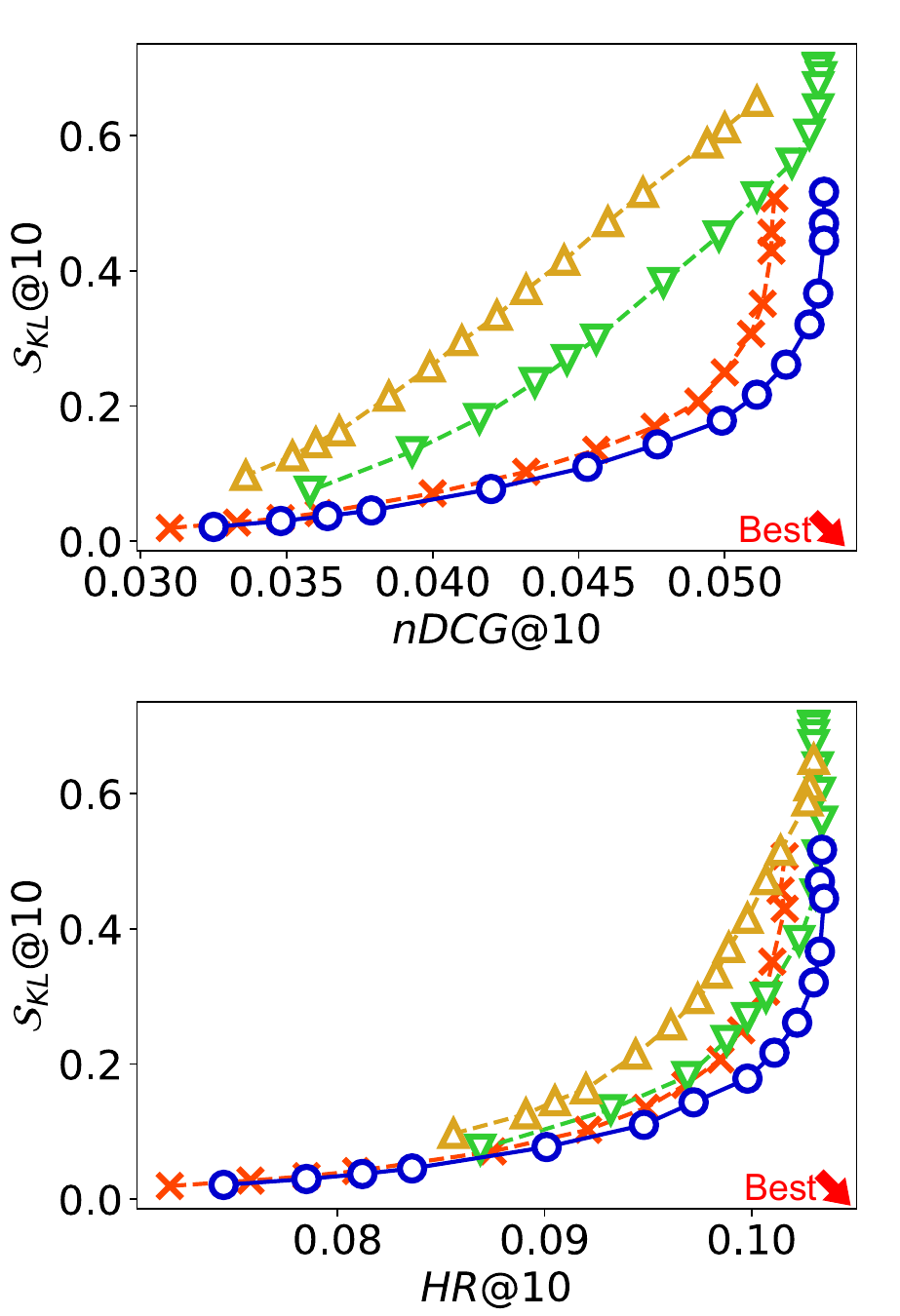}
                \vspace{-2em}
                \caption{Steam}
                \label{fig:ablation_steam}
        \end{subfigure}
        \caption{
        	Ablation study of \method.
        	The first row shows $nDCG@10$ vs. $\mathcal{S}_{KL}@10$, and the second row shows $HR@10$ vs. $\mathcal{S}_{KL}@10$.
            The main components of \method help improve the performance.
        }
        \label{fig:ablation}
\end{figure*}

\begin{figure}[t]
        \centering
        \begin{subfigure}{1.0\linewidth}
                \centering
                \includegraphics[width=0.7\linewidth]{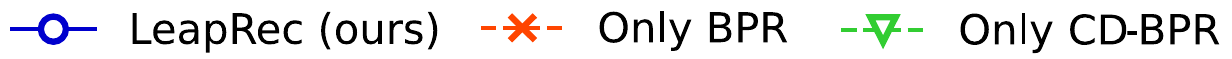}
                \label{fig:ablation2_legend}
                \vspace{-0.1em}
        \end{subfigure}
        \\
        \begin{subfigure}{1.0\linewidth}
                \centering
                \includegraphics[width=1\linewidth]{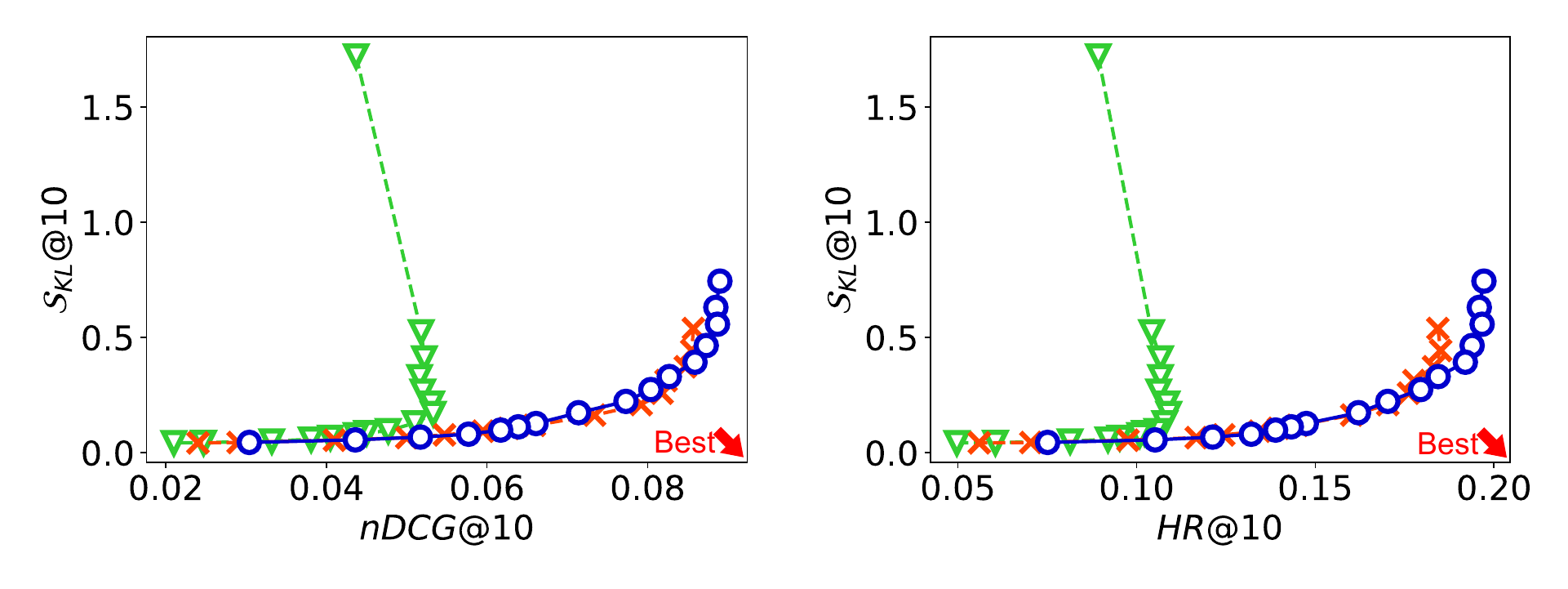}
                \vspace{-2.5em}
                \caption{ML-1M}
                \label{fig:ablation2_ml-1m}
        \end{subfigure}
        \\
        \vspace{-0.2em}
        \begin{subfigure}{1.0\linewidth}
                \centering
                \includegraphics[width=1\linewidth]{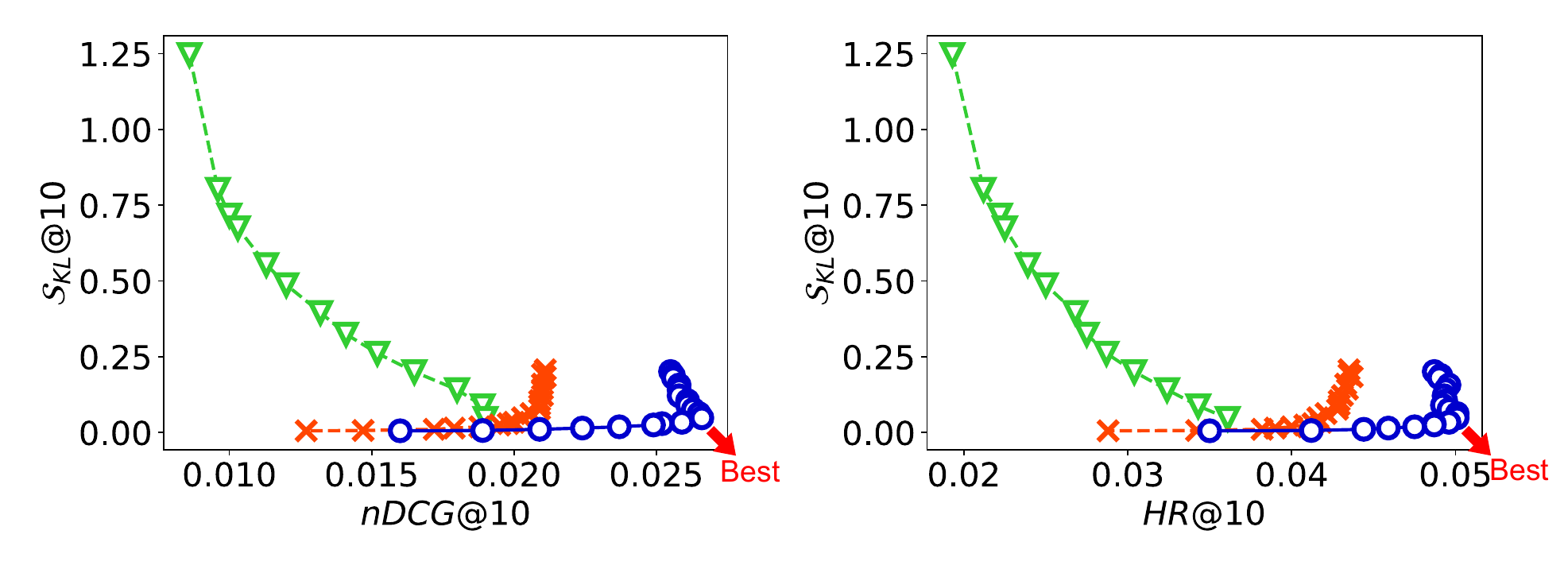}
                \vspace{-2.5em}
                \caption{Goodreads}
                \label{fig:ablation2_goodreads}
        \end{subfigure}
        \caption{
        	Investigation of calibration-disentangled learning-to-rank loss.
            Using only $\mathcal{L}_{\mathit{CD-BPR}}$ significantly reduces performance compared to using only $\mathcal{L}_{\mathit{BPR}}$ or a combined approach (\method), which demonstrates the importance of integrating both losses to achieve high accuracy and calibration.
        }
        \label{fig:ablation2}
\end{figure}

\vspace{-1em}
\subsubsection{Hyperparameters}
We implement \method using PyTorch~\cite{PaszkeGMLBCKLGA19}.
For all datasets, we set $\alpha$ (Equation~\eqref{eq:p}), $\beta$ (Equation~\eqref{eq:q_tilde}), and $\gamma$ (Equation~\eqref{eq:loss}) to $0.9$, $0.01$, and $0.1$, respectively.
For a backbone SASRec~\cite{KangM18}, we set the learning rate, the dimension of embedding (e.g., user embedding and item embedding), batch size, and the number of self-attention blocks to $0.001$, $50$, $128$, and $2$, respectively.
Moreover, the dropout rate is set to $0.2$ for ML-1M and $0.5$ for the other datasets.
In addition, the maximum sequence length is set to $200$ for ML-1M and $50$ for the other datasets.
Following previous work~\cite{SeymenAM21}, we use Gurobi software~\cite{gurobi21} in MIP, limiting the time to 10 minutes with an optimization gap $0.0001$ for every user separately.
We set the number of candidates to $1,000$ and $100$ respectively for MIP and MCF, following their settings.

\vspace{-1em}
\subsection{Performance Comparison (Q1)}
\label{subsec:performance_comparison}
\subsubsection{Trade-off Comparison}
\label{subsubsec:tradeoff}
In Figure~\ref{fig:overall}, we compare \method and baselines on four real-world datasets to verify whether \method provides better trade-off between accuracy and calibration than the baselines.
We present the performance of the backbone models for both \method (base) and SASRec (base), along with trade-off curves of the methods between accuracy and calibration.
The results show that \method consistently surpasses the baselines across all datasets, drawing better trade-off curves.
Notably, the observed higher accuracy of \method (base) compared to SASRec (base) shows that calibration-disentangled learning-to-rank offers an enhanced learning mechanism.
Additionally, \method maintains high accuracy levels while significantly enhancing calibration, unlike other competing methods.
Furthermore, both \method and CaliRec$^+$ achieve the lowest levels of sequential miscalibration by optimizing terms during reranking.
This advantage becomes particularly pronounced in datasets with longer user sequences, such as ML-1M and Goodreads, highlighting the importance of addressing sequential miscalibration as user interactions evolve over time.

We further expand our analysis by comparing the trade-off performance on ML-1M and Goodreads datasets using GRU4Rec\footnote{\url{https://github.com/jeon185/LeapRec/blob/main/experiments/GRU4Rec.pdf}}, Caser\footnote{\url{https://github.com/jeon185/LeapRec/blob/main/experiments/Caser.pdf}}, and BERT4Rec\footnote{\url{https://github.com/jeon185/LeapRec/blob/main/experiments/BERT4Rec.pdf}} as alternative backbone models.
Similar to the results presented in Figure~\ref{fig:overall}, \method outperforms the baselines significantly in terms of trade-off between accuracy and calibration, even when employing different backbone models such as GRU4Rec, Caser, and BERT4Rec.

\subsubsection{Complexity Comparison}
\label{subsubsec:complexity}
The model training phase of \method, which incorporates $\mathcal{L}_{\mathit{BPR}}$ and $\mathcal{L}_{\mathit{CD-BPR}}$, maintains a complexity level comparable to traditional methods.
Adding $\mathcal{L}_{\mathit{CD-BPR}}$ does not substantially increase the computational burden due to its efficient integration, with the Kullback-Leibler (KL) divergence computation being a $\mathcal{O}(d)$ task, where $d$ represents the dimensionality of embedding vectors.
The reranking phase is where \method distinctly differs from baselines in terms of time complexity.
In Table~\ref{table:complexity}, we compare the computational complexity of reranking algorithms.
The table shows that \method and similar greedy approaches like CaliRec and CaliRec$^+$ are notably efficient compared to the more complex algorithms such as MIP and MCF.

\subsubsection{Speed Comparison}
\label{subsubsec:speed}
In Table~\ref{table:time}, we compare the speed of \method and baselines by measuring the reranking times across four real-world datasets.
The batch size for all tests is standardized at 128 users.
For methods like MIP and MCF, $1,000$ candidate items per user are considered to ensure a fair comparison.
The results in the table confirm \method's competitive speed, comparable to CaliRec and CaliRec$^+$, thereby demonstrating its efficiency in rapidly generating recommendations.
Importantly, \method not only matches the speed of CaliRec and CaliRec$^+$ but also significantly surpasses baselines in achieving a superior trade-off between accuracy and calibration as described in Section~\ref{subsubsec:tradeoff}.

\subsection{Ablation Study (Q2)}
\label{subsec:ablation_study}
In Figure~\ref{fig:ablation}, we provide an ablation study that compares \method with its variants to evaluate the impact of its core components on performance.
The variant \emph{(-)CD} removes calibration-disentangled learning, by adopting a naive SASRec instead of our proposed training approach as the backbone.
The variants \emph{(-)PR-uniform} and \emph{(-)PR-reverse} remove Property~\ref{prop:relevance_priority} (i.e., relevance priority property) from the objective function in the reranking phase.
Specifically, \emph{(-)PR-uniform} applies a uniform balancing coefficient $\lambda$, diverging from the adaptive $\lambda^{1/k}$ used in Equation~\eqref{eq:objective}.
In contrast, \emph{(-)PR-reverse} inverses this adaptation by employing $\lambda^{k}$, prioritizing calibration in higher-ranked recommendations.
Hence, \emph{(-)PR-uniform} considers the relevance and calibration equally in all $k$'th recommendations, whereas \emph{(-)PR-reverse} prioritizes the calibration in the higher-ranked recommendations.
The results show the superiority of \method over its variants, verifying the effectiveness of its core components.
Notably, \method shows better performance than \emph{(-)CD} in most cases, indicating that disentangling calibration during the training of personalized rankings is essential for achieving both high accuracy and calibration.
In addition, \method consistently outperforms \emph{(-)PR-uniform} and \emph{(-)PR-reverse} in achieving better balances between accuracy and calibration across most cases.
Especially, \emph{(-)PR-reverse}, which deliberately inverts the relevance priority property, exhibits a more pronounced decline in performance compared to \emph{(-)PR-uniform}, which merely omits the property;
these observations validate the significance of entailing the relevance priority property in balancing accuracy and calibration.

In Figure~\ref{fig:ablation2}, we further explore the impact of exclusively using $\mathcal{L}_{\mathit{CD-BPR}}$, without using $\mathcal{L}_{\mathit{BPR}}$, on ML-1M and Goodreads datasets.
This analysis is presented separately due to the significantly different performance scales observed.
The variant \emph{Only BPR}, akin to \emph{(-)CD}, employs only $\mathcal{L}_{\mathit{BPR}}$ to train the backbone model.
In contrast, the variant \emph{Only CD-BPR} is trained solely with $\mathcal{L}_{\mathit{CD-BPR}}$.
The results show that \emph{Only CD-BPR} substantially underperforms compared to \method and \emph{Only BPR}.
This outcome supports the discussion in Section~\ref{subsec:learning} about the need to combine $\mathcal{L}_{\mathit{BPR}}$ and $\mathcal{L}_{\mathit{CD-BPR}}$ to achieve high performance both on accuracy and calibration.

\vspace{-1em}
\subsection{Effect of the balancing hyperparameter (Q3)}
\label{subsec:equity}
\begin{figure}[t]
	\centering
	\includegraphics[width=0.77\linewidth]{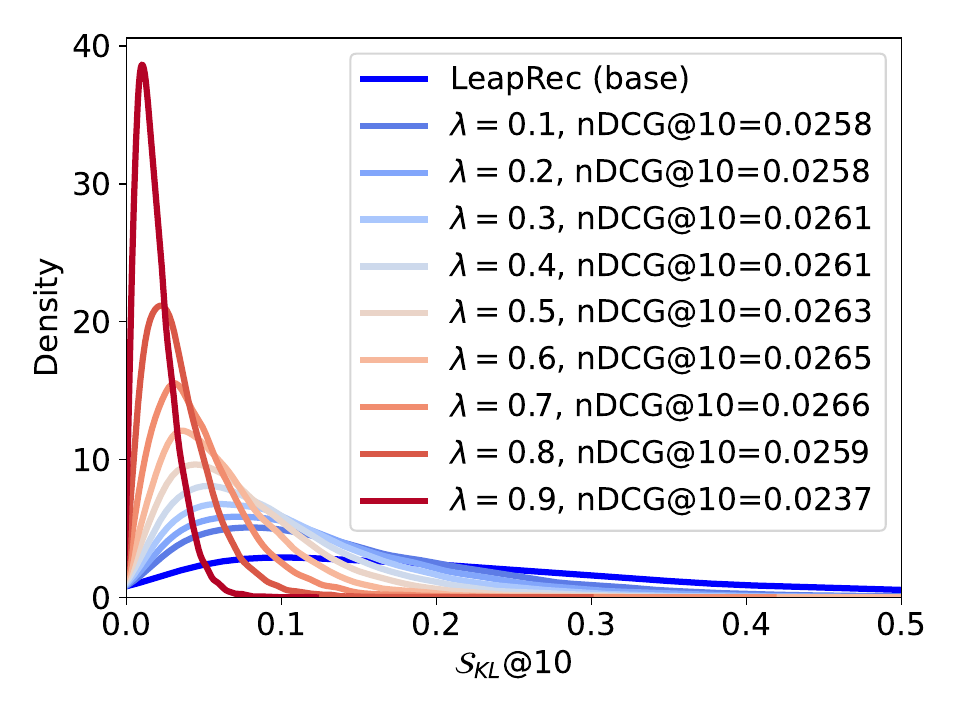}
    \vspace{-1em}
	\caption{
        Kernel density estimation (KDE) of sequential miscalibration $\mathcal{S}_{\mathit{KL}}@10$ on Goodreads dataset, while varying the balancing hyperparameter $\lambda$;
        range of the x-axis is set to $[0, 0.5]$.
        Averaged nDCG$@10$ for each setting is also written in the legend.
        The wide range of KDE indicates the recommendation quality in terms of calibration varies across users.
        \method successfully provides higher calibrated recommendations to more users as $\lambda$ increases.
    }
\label{fig:kdeplot}
\end{figure}

The balancing hyperparameter $\lambda$, as defined in Equation~\eqref{eq:objective}, is a key factor in controlling the balance between accuracy and calibration.
Increasing $\lambda$ leads to more calibrated recommendations at the cost of accuracy.
In Figure~\ref{fig:kdeplot}, we examine the impact of varying $\lambda$ on users' overall recommendation quality on Goodreads dataset.
We use kernel density estimation (KDE) to evaluate the distribution of sequential miscalibration $\mathcal{S}_{\mathit{KL}}@10$ across users under different $\lambda$.
In the figure, we also denote the average nDCG$@10$ to show the change in accuracy.
The results of low $\lambda$ (e.g., $0.1$) or missing calibration (\method (base)) show a wide range of distribution for $\mathcal{S}_{\mathit{KL}}@10$, meaning a substantial variance in calibration across users.
However, as $\lambda$ increases, recommendations become more uniformly calibrated across users.
The accuracy increases until $\lambda$ reaches $0.7$.

\begin{figure*}[t]
	\centering
	\includegraphics[width=0.929\linewidth]{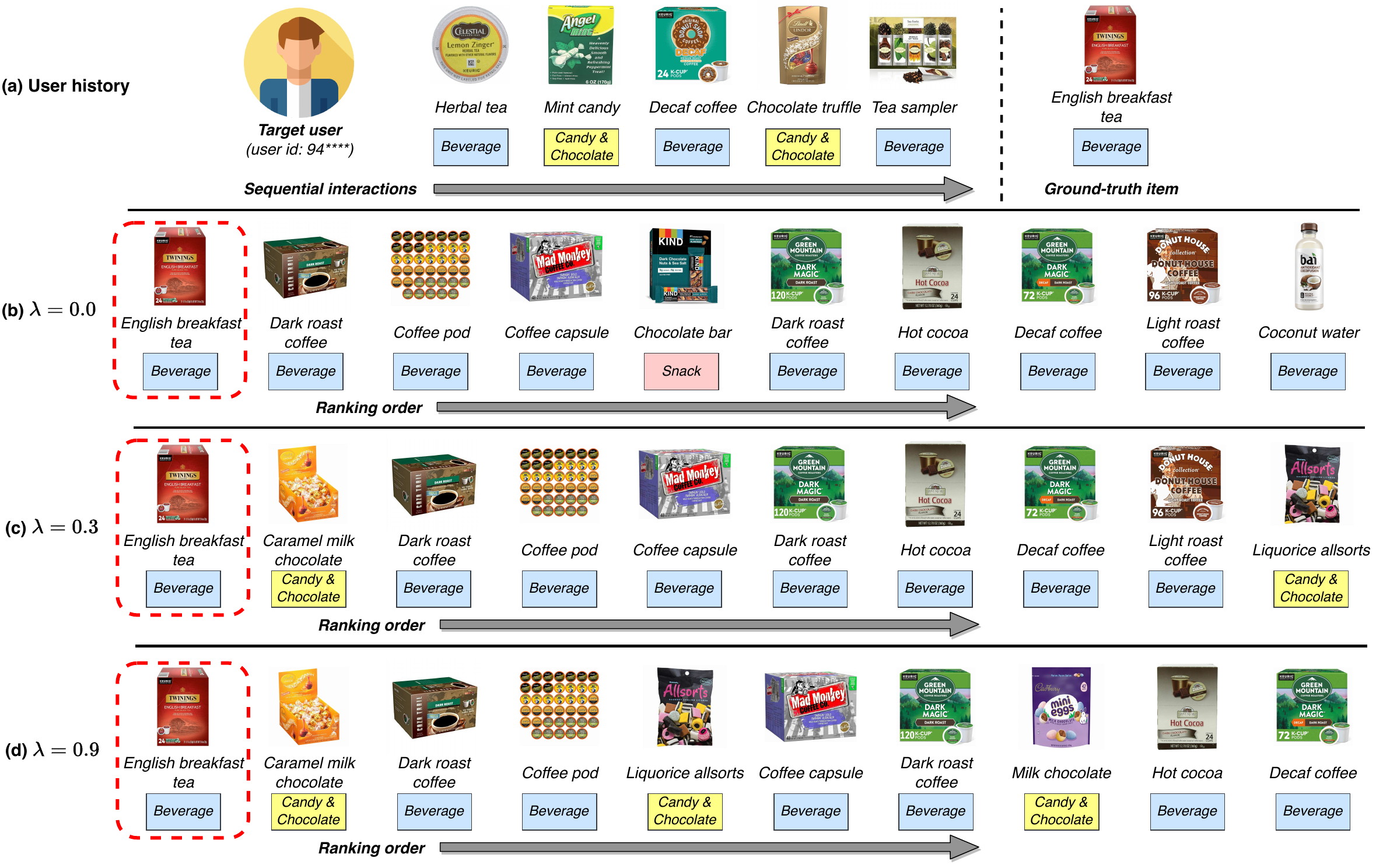}
	\caption{
        (a) A real user's sequential interactions and ground-truth next item in Grocery dataset.
        (b-d) The top-$10$ recommended items for the user by \method with different $\lambda$: $0.0$, $0.3$, and $0.9$.
        We mark the ground-truth item with a red box in each recommendation result.
        Colored boxes denote the categories.
        The values of $\mathcal{S}_{\mathit{KL}}@10$ are $1.5920$, $0.1001$, and $0.0212$ respectively for $\lambda=0.0$, $\lambda=0.3$, and $\lambda=0.9$.
        This case shows that \method enhances the calibration of the overall recommendation list while keeping relevant items at the top rank, as $\lambda$ increases.
    }
\label{fig:casestudy1}
\end{figure*}

\vspace{-1em}
\subsection{Case Study (Q4)} 
\label{subsec:qaulitative_analysis}
In Figure~\ref{fig:casestudy1}, we analyze a case to observe how \method balances relevance and calibration when recommending a list of items to a user on Grocery dataset.
We vary the calibration level for a random user and observe how the user experiences the recommendations.
Figure~\ref{fig:casestudy1} (a) represents the user's sequential interactions and the ground-truth item that the user will interact with.
Figures~\ref{fig:casestudy1} (b-d) show the recommendation results of \method for $\lambda$ (defined in Equation~\eqref{eq:objective}) values $0.0$, $0.3$, and $0.9$, respectively.
Figure~\ref{fig:casestudy1} (b) shows that \method accurately recommends the ground-truth item at rank $1$ without calibration (i.e., $\lambda=0.0$).
However, the overall items in the list skew towards the user's major interest \emph{Beverage}, narrowing the user experience for the recommendation.
In Figures~\ref{fig:casestudy1} (c-d), we observe that \method effectively considers the user's other interest \emph{Candy \& Chocolate} as well as the major interest \emph{Beverage} by enhancing calibration.
Notably, \method effectively retains \emph{English breakfast tea}, the most relevant item, at rank $1$ while improving calibration.
We further analyze another case where a user's category preference shifts towards a category previously not favored\footnote{\url{https://github.com/jeon185/LeapRec/blob/main/experiments/CaseStudy.pdf}}.

\section{Related Works}
\label{sec:related}
\boldheading{Calibrated recommendation.}
\citet{Steck18} first introduced calibrated recommendation to address the problem of traditional recommendation, where less dominant interests of users are often neglected.
The goal is to ensure that recommendation lists accurately reflect the proportion of categories users have shown interest in.
\citet{Steck18} proposed CaliRec which is a post-processing approach that greedily adjusts the recommended items to better match the user's historical category distribution.
Subsequent studies continued to adopt the post-processing strategy.
\citet{SeymenAM21} introduced a non-greedy approach by formulating calibration as a constrained optimization problem and solving it with a mixed integer programming (MIP) algorithm.
\citet{AbdollahpouriNG23} defined the calibrated recommendation as the maximum flow optimization problem and proposed a minimum cost flow (MCF) based algorithm.
On the other hand, \citet{chen2022dacsr} proposed DACSR, an end-to-end method, to simultaneously optimize accuracy and calibration in a single training phase.

Different from earlier advancements, our proposed method directly integrates calibration scores within the training phase, using a novel loss designed to enhance both the training and post-processing phases.
Additionally, we introduce a prioritized mechanism to effectively balance accuracy and calibration, addressing the dual criteria challenge more dynamically than previous methods.

\boldheading{Sequential recommendation.}
Sequential recommender systems predict users' future interactions by considering the temporal dynamics of their sequential interactions.
Such systems have shown their effectiveness by capturing the users' long-term and short-term interests~\cite{RendleFS10,HidasiKBT15}.
As early work, FPMC~\cite{RendleFS10} integrated first-order Markov chains with matrix factorization techniques~\cite{KorenBV09,SalakhutdinovM07} to simultaneously address users' sequential activities and overall preferences.
Subsequent research has evolved to include higher-order Markov chains~\cite{HeM16a, HeKM17}, capturing more complex sequential dependencies by considering multiple preceding interactions.
Within the last decade, deep learning models such as Recurrent Neural Networks (RNN)~\cite{HochreiterS97,ChoMGBBSB14}, Convolutional Neural Networks (CNN)~\cite{KrizhevskySH12}, and Transformers~\cite{VaswaniSPUJGKP17} have been adopted in sequential recommendation, marking significant progress through their ability to model non-linear relationships in user behaviors.
Models such as GRU4Rec~\cite{HidasiKBT15} and GRU4Rec$^+$~\cite{HidasiK18} demonstrated the effectiveness of GRU~\cite{ChoMGBBSB14} in session-based recommendations.
\citet{TangW18} effectively utilized CNN for extracting sequential patterns across both temporal and feature dimensions.
More recently, Transformer-based models such as SASRec~\cite{KangM18}, BERT4Rec~\cite{SunLWPLOJ19}, and SmartSense~\cite{JeonKYLK22} leveraged unidirectional and bidirectional Transformers to capture complex correlations within a sequence.
Other techniques include memory networks~\cite{huang2018improving, ChenXZTCQZ18}, translation learning~\cite{HeKM17}, hierarchical attention learning~\cite{ying2018sequential}, graph neural network learning~\cite{wu2019session, ma2020memory, chang2021sequential}, and contrastive learning~\cite{zhang2021causerec,ChenLLMX22,XieSLWGZDC22}.

We focus on integrating calibration within sequential recommender systems.
Our method is model-agnostic, allowing it to be applied across various sequential recommender systems.

\section{Conclusion}
\label{sec:conclusion}
In this work, we propose \method, a novel method that effectively balances accuracy and calibration in sequential recommendation.
\method first trains a backbone model using the proposed calibration-disentangled learning-to-rank loss to learn personalized rankings when calibration is considered.
Subsequently, \method applies the proposed relevance-prioritized reranking algorithm to the backbone's results, encouraging highly relevant items are placed at the top while accounting for calibration throughout the recommendations.
\method achieves superior performance over existing calibrated recommendation methods in extensive experiments.
Our ablation study further confirms the necessity of the core components of \method.
We also demonstrate through a case study how \method tackles relevance and calibration to achieve high performance on both.

\begin{acks}
{This work was supported by the National Research Foundation of Korea (NRF) grant funded by the Korea government (MSIT) (No. RS-2023-00244831).}
\end{acks}

\newpage

\bibliographystyle{ACM-Reference-Format}
\balance
\bibliography{paper}

\end{document}